\documentclass{aa}
\usepackage[fleqn]{amsmath}
\usepackage{graphicx,txfonts,color,upgreek}
\bibpunct{(}{)}{;}{a}{}{,}

\begin{document}
\title{The structure of disks around intermediate-mass young stars\\ from mid-infrared interferometry}
\subtitle{Evidence for a population of group II disks with gaps}
\author{J. Menu\inst{1,2}\fnmsep\thanks{PhD fellow of the Research Foundation -- Flanders (FWO)} \and R. van Boekel\inst{2} \and Th. Henning\inst{2} \and Ch. Leinert\inst{2} \and C. Waelkens\inst{1} \and L.~B.~F.~M.~Waters\inst{3,4}  
}

\institute{
Instituut voor Sterrenkunde, KU Leuven, Celestijnenlaan 200D, 3001 Leuven, Belgium \\ \email{jonathan.menu@ster.kuleuven.be} 
\and
Max Planck Institut f\"ur Astronomie, K\"onigstuhl 17, 69117 Heidelberg, Germany
\and
SRON Netherlands Institute for Space Research, Sorbonnelaan 2, 3584 CA Utrecht, The Netherlands
\and
Astronomical Institute Anton Pannekoek, University of Amsterdam, PO Box 94249, 1090 GE Amsterdam, The Netherlands
}

\date{Received / Accepted}

\authorrunning{J.~Menu et al.}
\titlerunning{The evolution of Herbig stars as seen with MIDI}

\abstract
{The disks around Herbig Ae/Be stars are commonly divided into group I and group II based on their far-infrared spectral energy distribution, and the common interpretation for that is flared and flat disks. Our understanding of the evolution of these disks is rapidly changing. Recent observations suggest that many flaring disks have gaps, whereas flat disks are thought to be gapless.}
{The different groups of objects can be expected to have different structural signatures in high-angular-resolution data, related to gaps, dust settling, and flaring. We aim to use such data to gain new insight into disk structure and evolution.}
{Over the past 10 years, the MIDI instrument on the Very Large Telescope Interferometer has collected observations of several tens of protoplanetary disks. We modeled the large set of observations with simple geometric models and compared the characteristic sizes among the different objects. A population of radiative-transfer models was synthesized for interpreting the mid-infrared signatures.}
{Objects with similar luminosities show very different disk sizes in the mid-infrared. This may point to an intrinsic diversity or could also hint at different evolutionary stages of the disks. Restricting this to the young objects of intermediate mass, we confirm that most group I disks are in agreement with being transitional (i.e., they have gaps). We find that several group II objects have mid-infrared sizes and colors that overlap with sources classified as group I, transition disks. This suggests that these sources have gaps, which has been demonstrated for a subset of them. This may point to an intermediate population between gapless and transition disks.}
{Flat disks with gaps are most likely descendants of flat disks without gaps. Potentially related to the formation of massive bodies, gaps may therefore even develop in disks in a far stage of grain growth and settling. The evolutionary implications of this new population could be twofold. Either gapped flat disks form a separate population of evolved disks or some of them may evolve further into flaring disks with large gaps. The latter transformation may be governed by the interaction with a massive planet, carving a large gap and dynamically exciting the grain population in the disk.}


\maketitle


\section{Introduction}\label{sect:intro}
Detailed study of the structure and evolution of protoplanetary disks is a necessary prerequisite for understanding planet formation. The dust- and gas-rich protoplanetary disks set the boundary conditions for the formation, initial dynamics, composition, and even the actual presence of planetary systems. Planetary systems are observed to be extremely diverse, so that understanding this diversity translates into constraining the onset of planet formation within the disks.

\begin{figure*}
 \centering
 \includegraphics[width=.8\textwidth,viewport=100 470 500 665,clip]{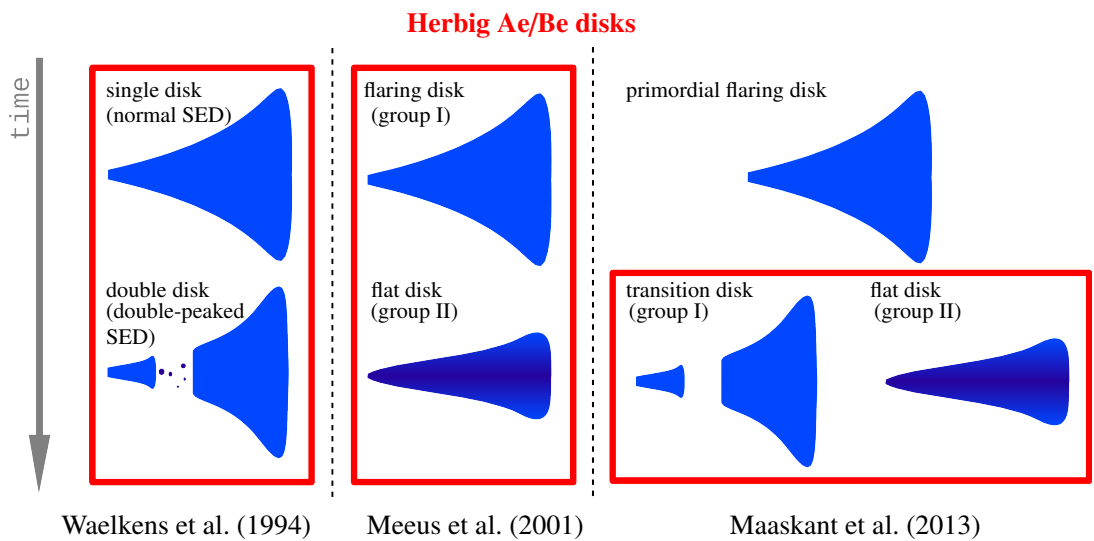}
 \caption{Different evolution scenarios proposed for Herbig Ae/Be disks. The Herbig Ae/Be phase is indicated in the red boxes. }\label{fig:scenarios}
\end{figure*}

Objects that have been under special scrutiny are the intermediate-mass young objects known as Herbig Ae/Be stars \citep{1960ApJS....4..337H,1998ARA&A..36..233W}. From the point of view of star formation, they present the link between the lower-mass T Tauri objects and the massive young stellar objects. From the point of view of planet formation, on the other hand, they represent the progenitors of debris-disk objects around A-type stars with or without detected planetary systems, such as Vega, $\beta$ Pic, Fomalhaut, and HR\,4796A \citep{1998Natur.392..788H}.\footnote{It is interesting to note that several of the few directly imaged planetary companions are found around A-type stars, which are descendants of Herbig Ae/Be stars. Examples of A-type exoplanet host stars are HR\,8799 (e.g., \citealt{2008Sci...322.1348M}), HD\,95086 (e.g., \citealt{2013ApJ...772L..15R}), $\kappa$\,And (e.g., \citealt{2013ApJ...763L..32C}), $\beta$\,Pic (e.g., \citealt{2010Sci...329...57L}), and HD\,100546 (e.g., \citealt{2013ApJ...766L...1Q}).}

Our understanding of the evolution of Herbig Ae/Be star disks (or, Herbig Ae/Be disks), itself, is in evolution. The idea that the disks are gradually dispersed and the central objects evolve into ``naked'' main-sequence stars is well established (e.g., \citealt{1987IAUS..115....1L}). Additionally, the potential mechanisms that drive the dispersion process have probably been identified (see, e.g., the overview in \citealt{2014prpl.conf..475A}). However, the coupling of these dissipation mechanisms to observational signatures for the global structure of the disks is highly non-trivial. We give an overview of the proposed evolutionary scenarios for Herbig Ae/Be disks in Fig.~\ref{fig:scenarios}, and describe them below.

\citet{1994ASPC...62..405W} have distinguished a class of Herbig Ae/Be stars with a broad dip in the 10-$\mu$m region of their infrared excess. This double-peaked spectral energy distribution (SED) was interpreted as representing a transition phase between classical broad infrared excesses of young stars and faint cool excesses of debris-disk objects. The conjectured explanation for this dip by these authors was the development of a physical gap in the radial dust distribution, which is also the interpretation of   \citet{1997A&A...324L..33V}. The origin of this gap would be the formation of larger bodies within the disk. The full evolutionary scenario that then emerged was shown in \citet{1998A&A...331..211M}.

A new classification of the Herbig Ae/Be objects has been proposed by \citet{2001A&A...365..476M}. Based on the shape of its mid- to far-infrared continuum, a source was classified as ``group II'' when the continuum was like a power law and  ``group I'' when an additional cold component was present on the power-law continuum. The main hypothesis for the spectral behavior is a morphological difference: group I sources having flaring disks, and group II sources have flat (or self-shadowed) disks. The physical origin for this difference could be grain growth \citep{2004A&A...417..159D}, decreasing the opacity throughout the disk, and/or grain settling \citep{2004A&A...421.1075D}, decreasing the irradiated surface of the disk. A likely evolutionary scenario for disks, where grains are expected to be growing and settling, is that group I sources evolve into group II sources. 

These evolutionary scenarios are based entirely on spatially unresolved observations. High-angular-resolution mid-infrared imaging of Herbig Ae/Be objects is challenging the observational picture. \citet{2012ApJ...752..143H} point out that many group I sources are found to have a gapped disk. A new evolutionary scenario proposed by \citet{2013A&A...555A..64M} is that group I and group II sources are different successors of a common ancestor: a primordial flaring disk. Gap formation would then have preceded the collapse of the outer disk in transitional group I sources, whereas in group II sources, grain growth and settling would have flattened the outer disk.

The notion that all group I sources may have gaps confirms the importance of spatially resolved observations. Typical radii of directly detected gaps are $\gtrsim20\,$au \citep{2013A&A...555A..64M}, which is close to the resolution limit of the observations (for a typical distance of 150\,pc). Moreover, for the group II sources, the amount of structural information from direct observations is very limited, since the disks tend to appear faint and small. An interesting alternative to direct imaging is mid-infrared interferometry:
\begin{enumerate}
 \item the angular resolution for typical observations is a factor of 10 higher than for direct imaging;
 \item the mid-infrared wavelength range corresponds to the thermal emission of small dust grains in the inner $1-10$\,au (scaling with the stellar luminosity, see \citealt{2011ppcd.book..114H}), which is a good tracer for the disk structure;
 \item spectrally resolved observations in the $8-13$\,micron atmospheric window provide additional information on the composition of the disk's small grain population.
\end{enumerate}
Mid-infrared interferometry was successfully used to resolve the gaps in the group I sources HD\,100546 ($\sim13\,$au; \citealt{2010A&A...511A..75B}), HD\,139614 ($\sim6$\,au; \citealt{2014A&A...561A..26M}), and HD\,179218 ($\sim10\,$au; Menu et al., in preparation).

Given this new evolutionary scenario for Herbig Ae/Be disks, many new questions need to be addressed. Are group I and II sources really two distinct classes? If a ``common ancestor'' for the two groups exists, what is its nature? Do both groups evolve into classical debris-disk objects and/or planetary systems? 

In this work, we aim at investigating the structural differences within Herbig Ae/Be disks, and protoplanetary disks in general, from a large set of mid-infrared interferometric data. During its ten years of operation, the MID-infrared Interferometric instrument (MIDI) on the Very Large Telescope Interferometer (VLTI) has been used to observe several dozens of protoplanetary disk objects. This work presents a statistically relevant compilation of targets observed with this instrument. In Sect.~\ref{sect:sample}, we give an overview of the sample, the observations, and the data reduction. Section \ref{sect:model} discusses the model choice for interpreting these data and the modeling results are presented in Sect.~\ref{sect:results}. In Sect.~\ref{sect:sizelum}, we present a size-luminosity relation for the full set of protoplanetary disks. This relation covers a broad range in stellar properties, which is interesting but also possibly limits the diagnostic power for structural differences within a subgroup. We limit the further analysis in Sect.~\ref{sect:sizecolor} to the Herbig Ae stars in the sample. Finally, we discuss the results for this group of stars in Sect.~\ref{sect:discussion}, in terms of the evolution of the disks. A summary and conclusions of this work can be found in Sect.~\ref{sect:conclusion}.

\begin{table*}
\centering
 \caption{Overview of the sample and relevant stellar properties. HAe/HBe = Herbig Ae/Be star, TT = T\,Tauri star, embHAe = embedded Herbig Ae star, DD = debris disk object.}\label{table:sample}
 {\tiny\begin{tabular}{rlccrcrcl}
 \hline\hline
 $\#$&name&R.A.~(J2000)&Dec.~(J2000)&type&$d$&$T_\mathrm{eff}$&$L$&references \\
 &&(h m s)&($^\circ$ $'$ $''$)&&(pc)&(K)&($L_\odot$)\\\hline

1	&	LkH$\alpha$\,330	&	$03\,45\,48.28$&$+32\,24\,11.9$	&	TT	&$	250	\pm	50	$&	5800	&	$11^{+5}_{-4}$	&	1,2,a	\\
2	&	V892\,Tau	&	$04\,18\,40.61$&$+28\,19\,15.5$	&	HAe	&$	142	\pm	14	$&	11220	&	$80^{+16}_{-40}$	&	3,b	\\
3	&	RY\,Tau	&	$04\,21\,57.41$&$+28\,26\,35.5$	&	TT/HAe	&$	142	\pm	14	$&	6310	&	$15^{+8}_{-6}$	&	3,b	\\
4	&	LkCa\,15	&	$04\,39\,17.79$&$+22\,21\,03.4$	&	TT	&$	142	\pm	14	$&	4350	&	$0.8^{+0.2}_{-0.2}$	&	3,c	\\
5	&	DR\,Tau	&	$04\,47\,06.20$&$+16\,58\,42.8$	&	TT	&$	142	\pm	14	$&	4060	&	$1.1^{+0.2}_{-0.2}$	&	3,c	\\
6	&	GM\,Aur	&	$04\,55\,10.98$&$+30\,21\,59.5$	&	TT	&$	142	\pm	14	$&	4730	&	$1.2^{+0.4}_{-0.4}$	&	3,c	\\
7	&	AB\,Aur	&	$04\,55\,45.84$&$+30\,33\,04.2$	&	HAe	&$	139	\pm	19	$&	9800	&	$60^{+26}_{-22}$	&	4,d	\\
8	&	SU\,Aur	&	$04\,55\,59.38$&$+30\,34\,01.5$	&	TT/HAe	&$	146	\pm	50	$&	5860	&	$11^{+8}_{-8}$	&	4,c	\\
9	&	HD\,31648	&	$04\,58\,46.26$&$+29\,50\,36.9$	&	HAe	&$	137	\pm	25	$&	8200	&	$23^{+9}_{-9}$	&	4,e	\\
10	&	UX\,Ori	&	$05\,04\,29.98$&$-03\,47\,14.2$	&	HAe	&$	460	\pm	50	$&	8710	&	$78^{+39}_{-35}$	&	5,6,f	\\
11	&	HD\,36112	&	$05\,30\,27.52$&$+25\,19\,57.0$	&	HAe	&$	279	\pm	70	$&	7800	&	$49^{+29}_{-28}$	&	4,e	\\
12	&	HD\,36917	&	$05\,34\,46.98$&$-05\,34\,14.5$	&	HAe	&$	375	\pm	30	$&	10000	&	$230^{+50}_{-50}$	&	7,e	\\
13	&	CQ\,Tau	&	$05\,35\,58.46$&$+24\,44\,54.0$	&	HAe	&$	113	\pm	23	$&	6750	&	$5.4^{+3.7}_{-2.9}$	&	4,b	\\
14	&	V1247\,Ori	&	$05\,38\,05.24$&$-01\,15\,21.6$	&	HAe	&$	385	\pm	15	$&	7250	&	$23^{+3}_{-3}$	&	8,f	\\
15	&	HD\,38120	&	$05\,43\,11.89$&$-04\,59\,49.8$	&	HAe	&$	375	\pm	30	$&	11000	&	$69^{+24}_{-21}$	&	7,e	\\
16	&	$\beta$\,Pic	&	$05\,47\,17.08$&$-51\,03\,59.4$	&	DD	&$	19.44	\pm	0.05	$&	8052	&	$8.6^{+0.8}_{-0.1}$	&	4,g	\\
17	&	HD\,45677	&	$06\,28\,17.42$&$-13\,03\,11.1$	&	HBe	&$	279	\pm	73	$&	21380	&	$600^{+460}_{-370}$	&	4,h	\\
18	&	VY\,Mon	&	$06\,31\,06.92$&$+10\,26\,04.9$	&	embHAe	&$	800	\pm	300	$&	12023	&	$3800^{+3300}_{-3000}$	&	9,10,i	\\
19	&	HD\,259431	&	$06\,33\,05.19$&$+10\,19\,19.9$	&	HAe	&$	660	\pm	100	$&	14000	&	$760^{+500}_{-350}$	&	7,e	\\
20	&	R\,Mon	&	$06\,39\,09.94$&$+08\,44\,09.7$	&	HBe	&$	760	\pm	300	$&	30903	&	$2500^{+4800}_{-2400}$	&	10,j	\\
21	&	HD\,50138	&	$06\,51\,33.39$&$-06\,57\,59.4$	&	HBe	&$	392	\pm	86	$&	15490	&	$2500^{+1200}_{-1200}$	&	4,h	\\
22	&	HD\,72106	&	$08\,29\,34.89$&$-38\,36\,21.1$	&	HAe	&$	279	\pm	88	$&	8750	&	$42^{+28}_{-28}$	&	4,k	\\
23	&	HD\,87643	&	$10\,04\,30.28$&$-58\,39\,52.0$	&	HBe	&$	1500	\pm	500	$&	17000	&	$41000^{+58000}_{-34000}$	&	11,l	\\
24	&	CR\,Cha	&	$10\,59\,06.97$&$-77\,01\,40.3$	&	TT	&$	160	\pm	15	$&	4900	&	$3.0^{+0.8}_{-0.7}$	&	12,m	\\
25	&	HD\,95881	&	$11\,01\,57.61$&$-71\,30\,48.3$	&	HAe	&$	170	\pm	30	$&	8990	&	$27^{+12}_{-11}$	&	13,n	\\
26	&	DI\,Cha	&	$11\,07\,20.72$&$-77\,38\,07.2$	&	TT	&$	160	\pm	15	$&	5860	&	$9.5^{+2.2}_{-2.2}$	&	12,m	\\
27	&	HD\,97048	&	$11\,08\,03.32$&$-77\,39\,17.4$	&	HAe	&$	158	\pm	16	$&	10000	&	$40^{+10}_{-10}$	&	4,h	\\
28	&	HP\,Cha	&	$11\,08\,15.09$&$-77\,33\,53.1$	&	TT	&$	160	\pm	15	$&	4205	&	$3.2^{+4.5}_{-1.5}$	&	12,m	\\
29	&	FM\,Cha	&	$11\,09\,53.40$&$-76\,34\,25.5$	&	TT	&$	160	\pm	15	$&	4350	&	$5.8^{+1.1}_{-2.3}$	&	12,m	\\
30	&	WW\,Cha	&	$11\,10\,00.10$&$-76\,34\,57.8$	&	TT	&$	160	\pm	15	$&	4350	&	$6.5^{+3.7}_{-2.3}$	&	12,m 	\\
31	&	CV\,Cha	&	$11\,12\,27.70$&$-76\,44\,22.3$	&	TT	&$	160	\pm	15	$&	5410	&	$4.6^{+1.1}_{-1.0}$	&	12,m	\\
32	&	HD\,98922	&	$11\,22\,31.67$&$-53\,22\,11.4$	&	HAe	&$	1150	\pm	515	$&	10500	&	$5400^{+5400}_{-5200}$	&	4,e	\\
33	&	HD\,100453	&	$11\,33\,05.57$&$-54\,19\,28.5$	&	HAe	&$	122	\pm	10	$&	7400	&	$14^{+2}_{-3}$	&	4,n	\\
34	&	HD\,100546	&	$11\,33\,25.44$&$-70\,11\,41.2$	&	HAe	&$	97	\pm	4	$&	10500	&	$24^{+6}_{-3}$	&	4,n	\\
35	&	T\,Cha	&	$11\,57\,13.55$&$-79\,21\,31.5$	&	TT	&$	108	\pm	9	$&	5890	&	$1.2^{+1.9}_{-0.7}$	&	14,h	\\
36	&	HD\,104237	&	$12\,00\,05.08$&$-78\,11\,34.5$	&	HAe	&$	115	\pm	5	$&	8410	&	$49^{+6}_{-7}$	&	4,n	\\
37	&	HD\,109085	&	$12\,32\,04.22$&$-16\,11\,45.6$	&	DD	&$	18.28	\pm	0.06	$&	6784	&	$5.1^{+0.6}_{-0.2}$	&	4,g	\\
38	&	DK\,Cha	&	$12\,53\,17.23$&$-77\,07\,10.7$	&	embHAe	&$	178	\pm	18	$&	7200	&	$19^{+4}_{-7}$	&	12,o	\\
39	&	HD\,135344\,B	&	$15\,15\,48.43$&$-37\,09\,16.0$	&	HAe	&$	142	\pm	27	$&	6750	&	$12^{+5}_{-5}$	&	15,e	\\
40	&	HD\,139614	&	$15\,40\,46.38$&$-42\,29\,53.5$	&	HAe	&$	142	\pm	27	$&	7600	&	$9.3^{+3.9}_{-3.7}$	&	15,d	\\
41	&	HD\,142666	&	$15\,56\,40.02$&$-22\,01\,40.0$	&	HAe	&$	140	\pm	20	$&	7900	&	$19^{+6}_{-6}$	&	7,16,e	\\

\end{tabular}
}
\end{table*}

\setcounter{table}{0}

\begin{table*}
\centering
 \caption{continued.}
 {\tiny\begin{tabular}{rlccrcrcl}
 \hline\hline
 $\#$&name&R.A.~(J2000)&Dec.~(J2000)&type&$d$&$T_\mathrm{eff}$&$L$&references \\
 &&(h m s)&($^\circ$ $'$ $''$)&&(pc)&(K)&($L_\odot$)\\\hline

42	&	HD\,142527	&	$15\,56\,41.88$&$-42\,19\,23.2$	&	HAe	&$	233	\pm	53	$&	6260	&	$50^{+25}_{-25}$	&	4,n	\\
43	&	HD\,142560	&	$15\,56\,42.31$&$-37\,49\,15.5$	&	TT	&$	150	\pm	30	$&	4000	&	$3.7^{+1.9}_{-1.5}$	&	17$\dag$,p	\\
44	&	HD\,143006	&	$15\,58\,36.91$&$-22\,57\,15.2$	&	TT	&$	145	\pm	15	$&	5884	&	$3.4^{+0.8}_{-0.7}$	&	18,19,20,q	\\
45	&	HD\,144432	&	$16\,06\,57.95$&$-27\,43\,09.7$	&	HAe	&$	160	\pm	29	$&	7500	&	$17^{+10}_{-7}$	&	4,e	\\
46	&	HD\,144668	&	$16\,08\,34.28$&$-39\,06\,18.3$	&	HAe	&$	142	\pm	27	$&	8200	&	$70^{+29}_{-29}$	&	15,e	\\
47	&	V2246\,Oph	&	$16\,26\,03.02$&$-24\,23\,36.0$	&	TT	&$	120	\pm	5	$&	5248	&	$7.5^{+2.0}_{-1.2}$	&	21,r	\\
48	&	HBC\,639	&	$16\,26\,23.35$&$-24\,20\,59.7$	&	TT	&$	120	\pm	5	$&	5250	&	$8.1^{+3.7}_{-3.3}$	&	21,s	\\
49	&	Elias\,2-24	&	$16\,26\,24.07$&$-24\,16\,13.5$	&	TT	&$	120	\pm	5	$&	4266	&	$1.8^{+0.2}_{-0.7}$	&	21,t	\\
50	&	Elias\,2-28	&	$16\,26\,58.44$&$-24\,45\,31.8$	&	TT	&$	120	\pm	5	$&	4169	&	$0.3^{+0.3}_{-0.1}$	&	21,t	\\
51	&	Elias\,2-30	&	$16\,27\,10.27$&$-24\,19\,12.7$	&	TT/HAe	&$	120	\pm	5	$&	5950	&	$6.3^{+5.6}_{-3.0}$	&	21,s	\\
52	&	V2129\,Oph	&	$16\,27\,40.27$&$-24\,22\,04.1$	&	TT	&$	120	\pm	5	$&	3981	&	$1.3^{+0.4}_{-0.3}$	&	21,t	\\
53	&	V2062\,Oph	&	$16\,31\,33.46$&$-24\,27\,37.2$	&	TT	&$	120	\pm	5	$&	4900	&	$1.5^{+0.6}_{-0.5}$	&	21,u	\\
54	&	HD\,150193	&	$16\,40\,17.92$&$-23\,53\,45.1$	&	HAe	&$	216	\pm	68	$&	9500	&	$110^{+70}_{-70}$	&	4,e	\\
55	&	AK\,Sco	&	$16\,54\,44.84$&$-36\,53\,18.5$	&	HAe	&$	103	\pm	21	$&	6500	&	$3.9^{+1.7}_{-1.7}$	&	4,v	\\
56	&	KK\,Oph	&	$17\,10\,08.13$&$-27\,15\,18.8$	&	HAe	&$	160	\pm	30	$&	8030	&	$6.2^{+2.9}_{-2.9}$	&	5,22,n	\\
57	&	51\,Oph	&	$17\,31\,24.95$&$-23\,57\,45.5$	&	HAe	&$	124	\pm	4	$&	10000	&	$240^{+25}_{-23}$	&	4,h	\\
58	&	HD\,163296	&	$17\,56\,21.28$&$-21\,57\,21.8$	&	HAe	&$	119	\pm	11	$&	9200	&	$38^{+10}_{-10}$	&	4,d	\\
59	&	HD\,169142	&	$18\,24\,29.77$&$-29\,46\,49.3$	&	HAe	&$	145	\pm	15	$&	7500	&	$11^{+4}_{-2}$	&	20,d	\\
60	&	MWC\,297	&	$18\,27\,39.52$&$-03\,49\,52.0$	&	HBe	&$	250	\pm	50	$&	25400	&	$21000^{+11000}_{-9000}$	&	23$\dag$,w	\\
61	&	MWC\,300	&	$18\,29\,25.69$&$-06\,04\,37.2$	&	HBe	&$	1800	\pm	200	$&	19000	&	$8300^{+8300}_{-4700}$	&	24,x	\\
62	&	R\,CrA	&	$19\,01\,53.65$&$-36\,57\,07.8$	&	HAe	&$	130	\pm	20	$&	11100	&	$90^{+150}_{-50}$	&	25,26,y	\\
63	&	T\,CrA	&	$19\,01\,58.77$&$-36\,57\,49.9$	&	HAe	&$	130	\pm	20	$&	6900	&	$4.7^{+4.1}_{-2.2}$	&	25,26,y	\\
64	&	HD\,179218	&	$19\,11\,11.25$&$+15\,47\,15.6$	&	HAe	&$	254	\pm	38	$&	9640	&	$100^{+30}_{-30}$	&	4,d	\\

\hline
 \end{tabular}}

\tablebib{{\tiny Distance references ($\dag=$ 20-$\%$ error assumed):
(1) \citet{2008ApJ...675L.109B};
(2) \citet{2006ApJ...646.1009K};
(3) \citet{1998MNRAS.301L..39W};
(4) \citet{2007A&A...474..653V};
(5) \citet{1992ApJ...397..613H};
(6) \citet{2000AJ....119.2919B};
(7) \citet{2013MNRAS.429.1001A};
(8) \citet{2013ApJ...768...80K};
(9) \citet{1994ApJ...436..807D};
(10) \citet{1997ApJ...489..210C};
(11) \citet{2009A&A...507..317M};
(12) \citet{1997A&A...327.1194W};
(13) \citet{2010A&A...516A..48V};
(14) \citet{2008hsf2.book..757T};
(15) \citet{2011A&A...530A..85M};
(16) \citet{2008hsf2.book..235P};
(17) \citet{2008hsf2.book..295C};
(18) \citet{1999AJ....117..354D};
(19) \citet{2008ApJ...683..479B};
(20) \citet{2012ApJ...752..143H};
(21) \citet{2008ApJ...675L..29L};
(22) \citet{2004A&A...423..537L};
(23) \citet{1997MNRAS.286..538D};
(24) \citet{2004A&A...417..731M};
(25) \citet{1981AJ.....86...62M};
(26) \citet{2003ApJ...584..853P}.
Effective-temperature references:
(a) \citet{2008ApJ...675L.109B};
(b) \citet{2004AJ....127.1682H};
(c) \citet{1995ApJS..101..117K};
(d) \citet{2012MNRAS.422.2072F};
(e) \citet{2013MNRAS.429.1001A};
(f) \citet{2002A&A...393..259M};
(g) \citet{2006AJ....132..161G};
(h) \citet{1998A&A...330..145V};
(i) \citet{1998A&AS..133...81T};
(j) \citet{1992ApJ...397..613H};
(k) \citet{2008MNRAS.391..901F};
(l) \citet{2009A&A...494..253K};
(m) \citet{2007ApJS..173..104L};
(n) \citet{2010ApJ...718..558A};
(o) \citet{2008ApJ...680.1295S};
(p) \citet{2007A&A...461..253S};
(q) \citet{2008ApJ...683..479B};
(r) \citet{2005AJ....129.2294M};
(s) \citet{2003ApJ...584..853P};
(t) \citet{2006A&A...452..245N};
(u) \citet{2013ApJ...769..149K};
(v) \citet{2003A&A...409.1037A};
(w) \citet{2008A&A...485..209A};
(x) \citet{2004A&A...417..731M};
(y) \citet{1992A&A...260..293B}.
}
}
\end{table*}

\section{MIDI sample}\label{sect:sample}
\subsection{MIDI and the interest of mid-infrared interferometry}
In 2002, MIDI \citep{2003Ap&SS.286...73L} was installed at the VLTI in Chile, and it became a unique instrument for its combination of high spatial resolution and spectral resolution ($R=30-230$) in the $N$ band ($\lambda=8-13\,\mu$m). For protoplanetary disks, MIDI is sensitive to the emission of small dust grains in the inner $1-10\,$au of the disk (assuming a typical distance of 150\,pc). For various reasons, this region is highly interesting and contains the imprint of ongoing physical processes. With MIDI, it has been shown that characteristic sizes of disks can be matched with the inner-disk geometry \citep{2004A&A...423..537L}. Another important finding is that the inner parts of protoplanetary disks can be highly crystalline \citep{2004Natur.432..479V}. Finally, the MIDI-detected geometry of the inner rim of disks can possibly be linked to the interaction with a planetary companion \citep{2013A&A...557A..68M,2014A&A...564A..93M}.

These discoveries show that MIDI provides a useful and necessary counterpart for high-spatial-resolution observations that are also sensitive to large grains (e.g., ALMA) or much hotter dust (e.g., PIONIER, AMBER).

\subsection{Sample and data overview}
Over the past ten years, over 100 young stellar objects with disks have been observed with MIDI. In this work, we focus on the intermediate-mass objects with, in general, low optical extinction. The objects are a compilation of sources from longer lists of (candidate) Herbig Ae/Be stars in \citet{1994A&AS..104..315T}, \citet{1998A&A...331..211M}, and \citet{2003AJ....126.2971V}. In total, data sets of 38 Herbig Ae/Be stars are included in the sample.

The Herbig Ae/Be stars represent only part of the young stellar objects (YSOs) that are observed with MIDI. \citet{2013A&A...558A..24B} present an overview and analysis of a large sample of massive YSOs observed with MIDI. The transition between the high-mass end of the Herbig Ae/Be stars and the genuine massive YSOs is not clear-cut, and we include a handful of probably more massive targets in the sample.

A third class of objects is the T\,Tauri stars. Full sample papers about MIDI observations for these objects are still underway. To extend our range in probed luminosities,  we included 22 representatives of the T\,Tauri class. The sources belong to classical low-mass star formation regions in Ophiuchus, Taurus, and Chamaeleon.

In Table \ref{table:sample}, an overview of the sample is given, including relevant stellar properties. An overview of the MIDI data that are used in this work is given in Appendix \ref{appendixB}. As can be seen from this table, the number of observations per target varies from 1 to 34 with a median of 6. In total, data obtained from about 240 nights are included in the paper. The data were obtained between 2003 and 2014, which is the full operational period of MIDI. A substantial amount of data are guaranteed time observations. Observations were done in different modes on both the 8.2-m Unit Telescopes (UTs) and 1.8-m Auxiliary Telescopes (ATs).

\subsection{Data reduction}
The large amount of observational data was reduced using the 2.0 version of the \texttt{EWS} software package \citep{2004SPIE.5491..715J}, released in October 2012. \texttt{EWS} is based on a coherent integration of the observed visibility signal. The data reduction consists of two principal steps: (1), the extraction of the raw data from the observations and (2) the calibration of these data. Step 1 is straightforward and is done observation by observation (for science targets and calibrators). Step 2, the calibration, requires the combination of science and calibrator data so is more involved. 

We analyzed the MIDI data in the form of correlated fluxes $F_\mathrm{corr}$, which we compared to the total fluxes $F$ of the disks (i.e., the spectra). Using correlated spectra is equivalent to using visibilities, which are the classical interferometric observables, but this has the advantage that the calibration does not invoke the photometric observations. The latter observations are often found to have low quality:
\begin{enumerate}
 \item The atmospheric and instrumental background contribution in the mid-infrared is strong and variable. Unlike correlated flux measurements, for which fringe scanning provides direct background subtraction at a high frequency (the frame rate, up to 160\,Hz), photometric observations require chopping, which is done at a much lower frequency (2\,Hz).
 \item The light path due to chopping differs for target and sky frames, and accordingly the background subtraction is more difficult. For the fringe measurement itself, the light path remains identical throughout the scans.
\end{enumerate}
In addition, for AT observations, the primary beam almost fills the entire instrumental field of view, which complicates the estimation of the sky signal next to the source, on the detector. This can make AT photometry unusable, even for sources as bright as 50\,Jy.

The calibrated correlated flux $F_{\mathrm{corr}}$ is calculated from the raw correlated flux $C_\mathrm{corr}$, as follows:
\begin{equation}
 F_{\mathrm{corr},\nu}=\frac{C_{\mathrm{corr},\nu}}{T^c_{\mathrm{corr},\nu}}, \qquad\textrm{where}\qquad T^c_{\mathrm{corr},\nu}=\frac{C^c_{\mathrm{corr},\nu}}{F^c_\nu\,V^c_\nu}. \label{eq:cal}
\end{equation}
(The index $c$ denotes calibrator quantities.) Here, $V^c$ is the calibrator visibility, as calculated from its known apparent diameter, and $F^c$ is the known calibrator spectrum. The total flux $F$ of a science target is calibrated in the same way as in Eq.~(\ref{eq:cal}) with $V^c\equiv1$. Each observation therefore leads to two observables: a (baseline-dependent) correlated flux measurement $F_\mathrm{corr}$ and a total flux measurement $F$. Finally, for a collection of observations of the same target, one can average the total flux measurements to a single high-quality spectrum. Indeed, the total spectrum is the equivalent of the correlated flux at zero baseline.

The variable atmospheric transmission and coherence losses make the calibration in Eq.~(\ref{eq:cal}) non-trivial: $T^c_\mathrm{corr}$, called the transfer function, is a time- and airmass-dependent quantity. We correct for the time dependency by doing a linear interpolation in time of the transfer function, calculated for different calibrators observed the same night. The airmass dependency is corrected for by fitting a line to the $\log T^c_\mathrm{corr}$ vs.~airmass diagram for the entire data set and applying the corresponding airmass-correction factor to the observations (see \citealt{2005A&A...437..189V} and \citealt{2012SPIE.8445E..1GB}). The uncertainty on $T^c_\mathrm{corr}$ includes both the intrinsic uncertainty of $T^c_\mathrm{corr}$ (photon noise) and the inter-calibrator variations of $T^c_\mathrm{corr}$ (airmass and temporal variations, evaluated as the standard deviation of all transfer-function observations).

All data presented in this paper will be made available in the Optical interferometry DataBase (OiDB), managed by the Jean-Marie Mariotti Center (JMMC).\footnote{\texttt{http://oidb.jmmc.fr/}}

\section{Model geometry}\label{sect:model}
The large difference in number of observations (i.e., UV points) per target makes the model choice difficult. Generally, two options can be considered for modeling the disk geometry from interferometric data: (1), a geometric model with a prescribed intensity distribution for the disk emission, and (2), a radiative-transfer model with a self-consistently calculated intensity distribution. Given the large sample, computational reasons make the second approach cumbersome, and we restrict ourselves to a geometric model.

In the near-infrared, inclined ring models have been successfully used for fitting the interferometric data of disks and determining their orientation (e.g., \citealt{2003ApJ...588..360E,2004ApJ...613.1049E, 2011A&A...531A..84B,2013A&A...551A..21K}). The situation in the mid-infrared is different. Unlike in the near-infrared, where essentially all disk emission comes from a compact region around the dust sublimation radius, the mid-infrared intensity distribution corresponds to a relatively extended region at different temperatures. Simple geometric models (e.g., rings, uniform disks, Gaussians) do not represent this complexity properly. A different but related problem is that the estimation of disk orientations tends to be difficult. Examples can be found in the literature: \citet{2008ApJ...676..490K} indicate that their disk orientation derived from near- and mid-infrared interferometry disagree; \citet{2008A&A...491..809F} find a disk orientation for HD\,135344\,B that differs substantially from more precise estimates \citep{2011AJ....142..151L} and their orientation for HD\,101412 ($i=80\pm7^\circ$) is rather unlikely for an unobscured Herbig star.

We propose the following, semi-physical model for the mid-infrared emission of the disks. The mid-infrared intensity distribution is assumed to come from an (vertically) optically thin surface layer of the disk:
\begin{equation}
 I_{\mathrm{disk},\nu}(R)=\tau_\nu\,B_\nu\big(T(R)\big),\quad\textrm{where}\quad T(R)=T_\mathrm{sub}\left(\frac{R}{R_\mathrm{sub}}\right)^{-q}.\label{eq:BBdisk}
\end{equation}
In this equation, $T_\mathrm{sub}$ and $R_\mathrm{sub}$ are the sublimation temperature and radius of the dust, physically corresponding to the inner rim of the disk. The radial range of the model is $R_\mathrm{sub}\leq R \leq R_\mathrm{out}$. The model has two free parameters: a constant optical depth $\tau_\nu$ and the temperature gradient $q$. Other parameters are fixed or precalculated: $T_\mathrm{sub} = 1500\,$K, $R_\mathrm{out}=300\,$au, and 
\begin{equation}
 R_\mathrm{sub}=R_\star\left(\frac{T_\mathrm{eff}}{T_\mathrm{sub}}\right)^2=\left(\frac {L_\star}{4\pi\sigma T_\mathrm{sub}^4}\right)^{1/2}\label{eq:rsub}
\end{equation}
(see, e.g., \citealt{2010ARA&A..48..205D}; note that this expression assumes that the rim itself is optically thick in the radial direction), where $R_\star$, $T_\mathrm{eff}$, and $L_\star$ are the stellar radius, effective temperature, and luminosity, respectively. 

We assume a pole-on orientation for all disks, similar to other interferometric surveys based on a small amount of data per target (e.g., \citealt{2002ApJ...579..694M,2005ApJ...624..832M}). On the one hand, this assumption allows us to model sources with only a few observations (i.e., where the UV coverage intrinsically does not allow determining the orientation). On the other hand, we prefer taking the same orientation for all objects rather than fitting orientations based on the data themselves.\footnote{As already mentioned, constraining disk orientations from mid-infrared interferometry turns out to be difficult. The main reason for this is the typically low and/or non-uniform (often unidirectional) UV sampling of the interferometric observations, in combination with an analytic description of the disk's brightness. Interestingly, when sharp radial edges are present in the resolution range of MIDI, strong constraints on the disk orientation are imprinted in the visibilities, providing the necessary diagnostics to fit this orientation (e.g., \citealt{2014A&A...561A..26M}; \citealt{2015A&A...578A..40H}; Menu et al., in preparation). For continuous disks, a good azimuthal sampling is a necessary prerequisite. In Sect.~\ref{sect:incbase}, it is shown that our approximation of a pole-on orientation has no major influence on our results.} Since most disks have mid-infrared emission features and not absorption features, an orientation close to edge-on is also very unlikely. Under the approximation of a pole-on orientation, the correlated flux of the disk measured at the wavelength $\lambda=c/\nu$ on a projected baseline length $B$ becomes
\begin{eqnarray}
 F_{\mathrm{corr.\,disk,}\nu}(B)=F_{\mathrm{disk,}\nu}\,\frac{\int_{R_\mathrm{sub}}^{R_\mathrm{out}}\mathrm dR\, R\,B_\nu\big(T(R)\big)\,J_0\big(2\pi R\,B/(d\,\lambda)\big)}{\int_{R_\mathrm{sub}}^{R_\mathrm{out}}\mathrm dR\, R\,B_\nu\big(T(R)\big)},
\end{eqnarray}
where $d$ is the distance, and $J_0$ the 0-th order Bessel function. The free parameter $\tau_\nu$ in Eq.~(\ref{eq:BBdisk}) is now absorbed in the free parameter $F_{\mathrm{disk,}\nu}$, the total disk flux at the given wavelength. The correlated flux $F_{\mathrm{corr,}\nu}$ for an observation is then simply (the absolute value of) the sum of $F_{\mathrm{corr.\,disk,}\nu}$ (the correlated flux of the disk) and $F_{\star,\nu}$ (the unresolved stellar flux).

For all objects, our model is fit to the data at $\lambda=10.7\,\mu$m, the reference wavelength of other mid-infrared high-angular-resolution disk surveys \citep{2009ApJ...700..491M,2013A&A...558A..24B}. The algorithm of \citet{2013PASP..125..306F} is used for the parameter estimation.

\section{Results}\label{sect:results}
In Fig.~\ref{fig:model}, the resulting fits of the temperature-gradient model to the data are shown per target. The corresponding parameters of the models are shown in Table \ref{table:fitpar}. In general, the temperature-gradient model in Eq.~(\ref{eq:BBdisk}) leads to a good qualitative and quantitative reproduction of the observed trends. The majority of the objects indicate a smoothly declining correlated-flux profile, a behavior that is captured well by the model. For a minority of objects, the resulting fits are rather poor, and we discuss possible limitations of the model below.

\begin{figure*}
\centering
 \includegraphics[width=.95\textwidth,viewport=10 70 1450 1980,clip]{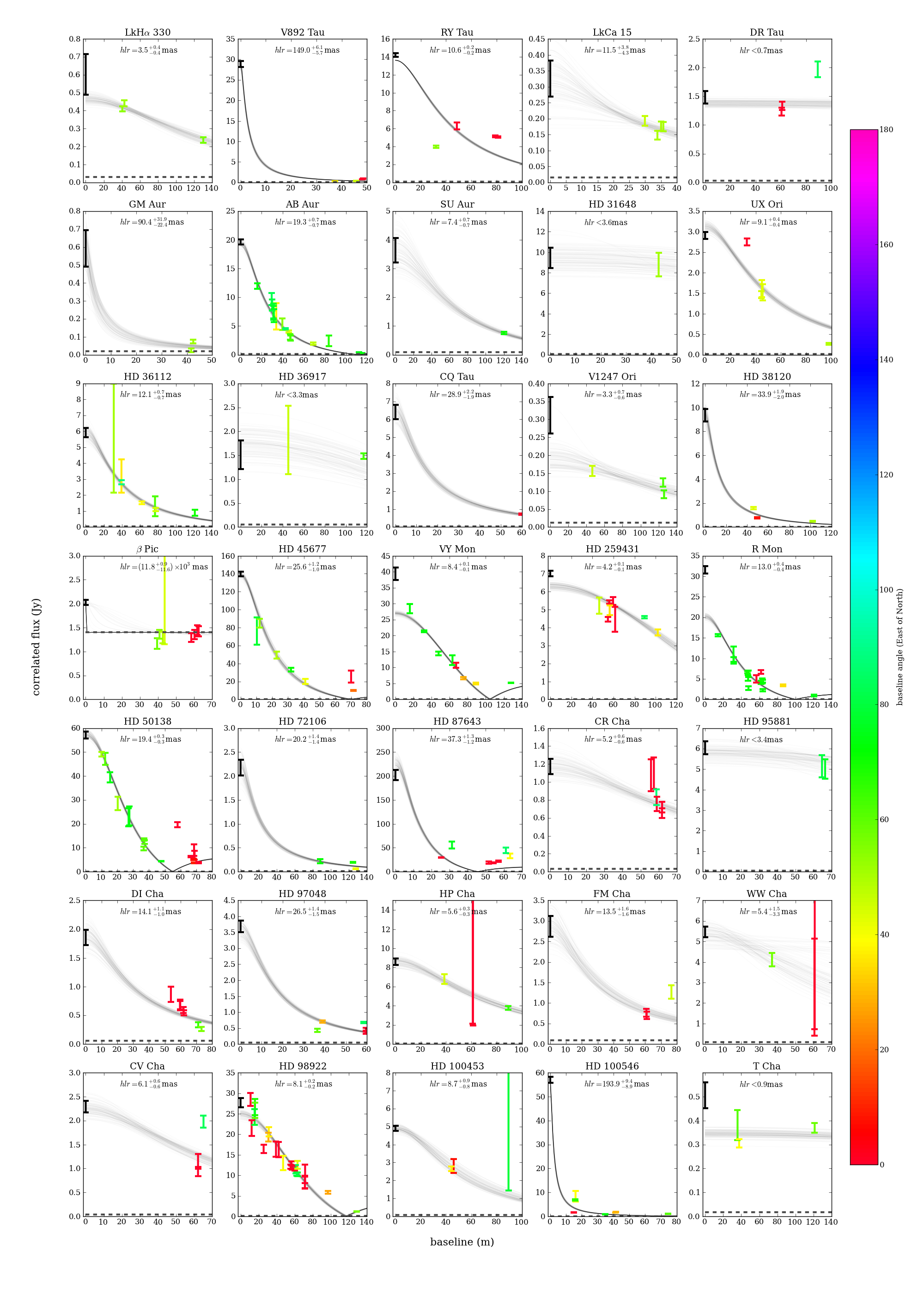}
 \caption{Fits of the temperature-gradient model to the 10.7\,$\mu$m correlated fluxes. The gray lines represent a range of possible models for the data, and the dashed line is the unresolved stellar contribution at $10.7\,\mu$m. The half-light radius $hlr$ (in mas) corresponding to the models is included in the plot.}\label{fig:model}
\end{figure*}

\setcounter{figure}{1}
\begin{figure*}
\centering
 \includegraphics[width=.95\textwidth,viewport=10 60 1450 1700,clip]{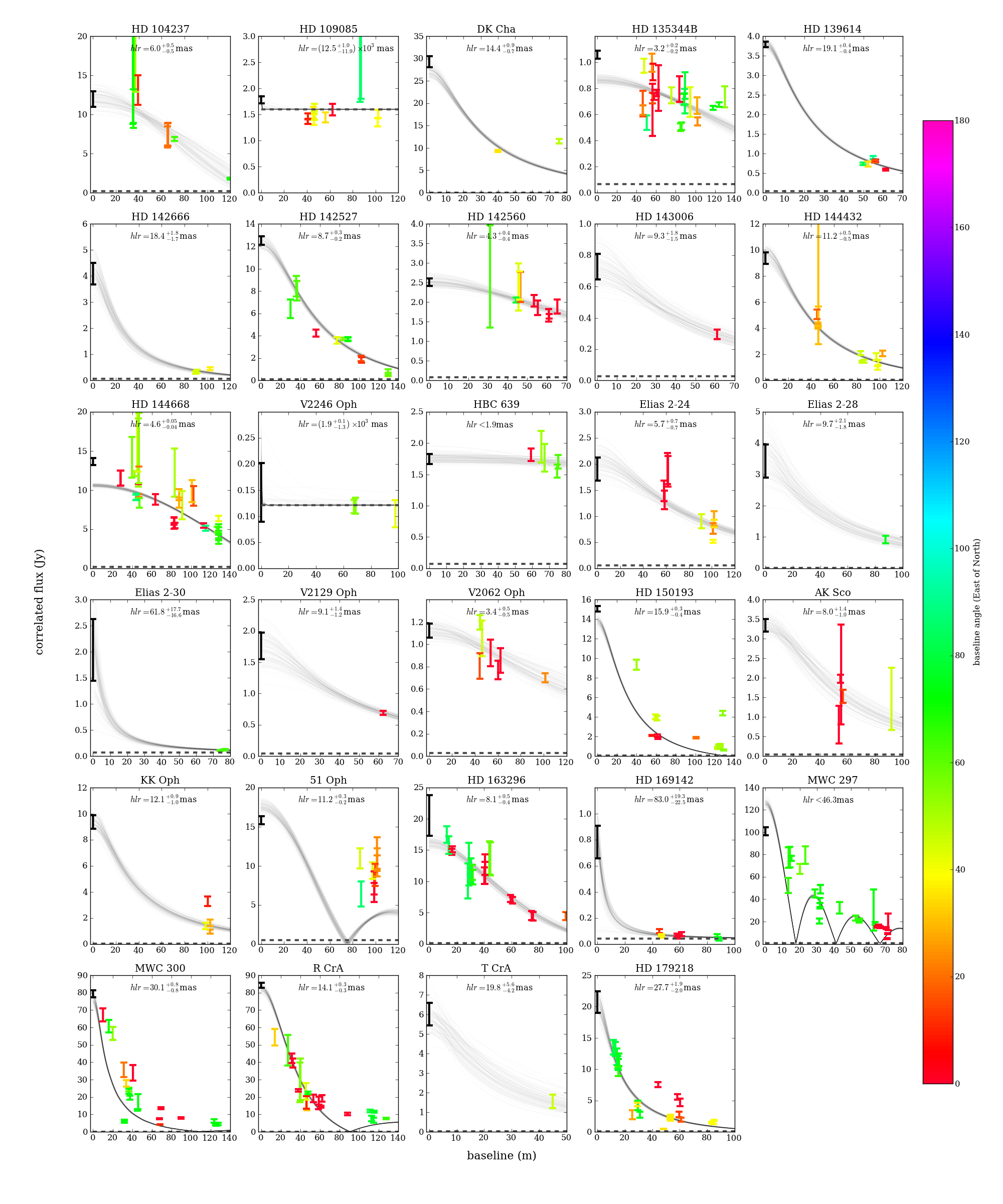}
 \caption{continued.}
\end{figure*}

First, we can consider the approximation that disks are oriented pole-on. Several objects with multiple observations along different baseline orientations indicate that the disk orientation indeed seems not too far from pole-on. Examples are HD\,98922, VY\,Mon, and HD\,142527. For other objects, such as HD\,50138, RY\,Tau, and UX\,Ori, we see clear hints of different geometric extents along different baseline angles. A likely explanation then is that we see these disks under a significant inclination. For the first two of these objects, the inclination is indeed observed to be relatively high ($i=56\pm4^\circ$ and $66\pm2^\circ$, see \citealt{2011A&A...528A..20B,2010ApJ...714.1746I}), and the inclination is also expected to be high for UX\,Ori \citep{1994A&A...292..165G}. Still, it is comforting to see that objects like HD\,142666 and HD\,144432, which are physically similar but probably have significantly different inclinations \citep{1991isrs.conf..345B}, are found to have similar extents. This suggests that the assumption of a pole-on orientation has no major influence on the outcome (discussed further in Sect.~\ref{sect:incbase}).

\begin{table*}
 \centering
 
 \caption{Fit parameters $F_{\mathrm{disk},\nu}$ and $q$ for the temperature gradient model, per object, and the derived half-light radii $hlr$ (in mas and au; the latter value involves the distance uncertainty) at $\lambda=10.7\,\mu$m. Error bars denote the 68-$\%$ confidence intervals.}\label{table:fitpar}
 {\tiny
 \begin{tabular}{rlcccc|rlcccc}
 \hline\hline $\#$&name&$F_{\mathrm{disk},\nu}$&$q$&$hlr$&$hlr$&$\#$&name&$F_{\mathrm{disk},\nu}$&$q$&$hlr$&$hlr$\\
 &&(Jy)&&(mas)&(au)&&&(Jy)&&(mas)&(au)\\\hline
 
1	&	LkH$\alpha$\,330	&	$0.44^{+0.02}_{-0.02}$	&	$0.73^{+0.03}_{-0.03}$	&	$3.5^{+0.4}_{-0.4}$	&	$0.9^{+0.2}_{-0.2}$	&	33	&	HD\,100453	&	$4.8^{+0.2}_{-0.1}$	&	$0.71^{+0.03}_{-0.03}$	&	$8.7^{+0.9}_{-0.8}$	&	$1.1^{+0.1}_{-0.1}$	\\
2	&	V892\,Tau	&	$28.8^{+0.6}_{-0.7}$	&	$0.435^{+0.003}_{-0.003}$	&	$149.0^{+6.1}_{-5.7}$	&	$21.2^{+2.3}_{-2.2}$	&	34	&	HD\,100546	&	$57.1^{+1.2}_{-1.3}$	&	$0.404^{+0.003}_{-0.003}$	&	$193.9^{+9.4}_{-8.9}$	&	$18.8^{+1.2}_{-1.2}$	\\
3	&	RY\,Tau	&	$13.5^{+0.02}_{-0.01}$	&	$0.642^{+0.003}_{-0.003}$	&	$10.6^{+0.2}_{-0.2}$	&	$1.5^{+0.2}_{-0.2}$	&	35	&	T\,Cha	&	$0.33^{+0.01}_{-0.01}$	&	$>2.0$	&	$<0.9$	&	$<0.10$	\\
4	&	LkCa\,15	&	$0.29^{+0.07}_{-0.07}$	&	$0.45^{+0.04}_{-0.02}$	&	$11.5^{+3.8}_{-4.3}$	&	$1.6^{+0.6}_{-0.6}$	&	36	&	HD\,104237	&	$11.9^{+0.8}_{-0.8}$	&	$1.5^{+0.3}_{-0.2}$	&	$6.0^{+0.5}_{-0.5}$	&	$0.69^{+0.07}_{-0.07}$	\\
5	&	DR\,Tau	&	$1.35^{+0.05}_{-0.05}$	&	$>1.6$	&	$<0.7$	&	$<0.10$	&	37	&	HD\,109085	&	$0.19^{+0.06}_{-0.07}$	&	$-0.5^{+0.9}_{-1.0}$	&	$(12.5^{+1.0}_{-11.9})$	&	$228.2^{+18.3}_{-217.4}$	\\
6	&	GM\,Aur	&	$0.6^{+0.1}_{-0.1}$	&	$0.35^{+0.01}_{-0.01}$	&	$90.4^{+31.9}_{-22.4}$	&	$12.8^{+4.7}_{-3.4}$	&	38	&	DK\,Cha	&	$27.3^{+1.2}_{-1.3}$	&	$0.57^{+0.01}_{-0.01}$	&	$14.4^{+0.9}_{-0.7}$	&	$2.6^{+0.3}_{-0.3}$	\\
7	&	AB\,Aur	&	$19.3^{+0.5}_{-0.4}$	&	$0.67^{+0.01}_{-0.01}$	&	$19.3^{+0.7}_{-0.7}$	&	$2.7^{+0.4}_{-0.4}$	&	39	&	HD\,135344\,B	&	$0.80^{+0.03}_{-0.02}$	&	$1.10^{+0.06}_{-0.05}$	&	$3.2^{+0.2}_{-0.2}$	&	$0.45^{+0.09}_{-0.09}$	\\
8	&	SU\,Aur	&	$3.5^{+0.5}_{-0.4}$	&	$0.68^{+0.02}_{-0.02}$	&	$7.4^{+0.7}_{-0.7}$	&	$1.1^{+0.4}_{-0.4}$	&	40	&	HD\,139614	&	$3.8^{+0.1}_{-0.1}$	&	$0.522^{+0.003}_{-0.002}$	&	$19.1^{+0.4}_{-0.4}$	&	$2.7^{+0.5}_{-0.5}$	\\
9	&	HD\,31648	&	$9.4^{+0.8}_{-0.8}$	&	$>1.4$	&	$<3.6$	&	$<0.5$	&	41	&	HD\,142666	&	$4.0^{+0.5}_{-0.4}$	&	$0.57^{+0.01}_{-0.01}$	&	$18.4^{+1.8}_{-1.7}$	&	$2.6^{+0.5}_{-0.4}$	\\
10	&	UX\,Ori	&	$3.1^{+0.1}_{-0.1}$	&	$0.61^{+0.01}_{-0.01}$	&	$9.1^{+0.4}_{-0.4}$	&	$4.2^{+0.5}_{-0.5}$	&	42	&	HD\,142527	&	$12.2^{+0.4}_{-0.4}$	&	$0.71^{+0.01}_{-0.01}$	&	$8.7^{+0.3}_{-0.2}$	&	$2.0^{+0.5}_{-0.5}$	\\
11	&	HD\,36112	&	$5.9^{+0.3}_{-0.3}$	&	$0.60^{+0.01}_{-0.01}$	&	$12.1^{+0.7}_{-0.7}$	&	$3.4^{+0.9}_{-0.9}$	&	43	&	HD\,142560	&	$2.4^{+0.1}_{-0.1}$	&	$0.67^{+0.02}_{-0.02}$	&	$4.3^{+0.4}_{-0.4}$	&	$0.65^{+0.15}_{-0.14}$	\\
12	&	HD\,36917	&	$1.7^{+0.2}_{-0.2}$	&	$>2.4$	&	$<3.3$	&	$<1.3$	&	44	&	HD\,143006	&	$0.69^{+0.08}_{-0.08}$	&	$0.55^{+0.02}_{-0.02}$	&	$9.3^{+1.8}_{-1.5}$	&	$1.3^{+0.3}_{-0.3}$	\\
13	&	CQ\,Tau	&	$6.4^{+0.4}_{-0.4}$	&	$0.48^{+0.01}_{-0.01}$	&	$28.9^{+2.2}_{-1.9}$	&	$3.3^{+0.7}_{-0.7}$	&	45	&	HD\,144432	&	$9.6^{+0.4}_{-0.5}$	&	$0.62^{+0.01}_{-0.01}$	&	$11.2^{+0.5}_{-0.5}$	&	$1.8^{+0.3}_{-0.3}$	\\
14	&	V1247\,Ori	&	$0.18^{+0.03}_{-0.02}$	&	$0.73^{+0.06}_{-0.05}$	&	$3.3^{+0.7}_{-0.6}$	&	$1.3^{+0.3}_{-0.2}$	&	46	&	HD\,144668	&	$10.4^{+0.2}_{-0.1}$	&	$3.6^{+0.3}_{-0.2}$	&	$4.55^{+0.05}_{-0.04}$	&	$0.65^{+0.12}_{-0.12}$	\\
15	&	HD\,38120	&	$9.3^{+0.5}_{-0.5}$	&	$0.47^{+0.01}_{-0.01}$	&	$33.9^{+1.9}_{-2.0}$	&	$12.7^{+1.2}_{-1.3}$	&	47	&	V2246\,Oph	&	$0.05^{+0.04}_{-0.04}$	&	$-0.6^{+0.9}_{-1.0}$	&	$(1.9^{+0.1}_{-1.3})$	&	$229.6^{+19.9}_{-160.6}$	\\
16	&	$\beta$\,Pic	&	$0.62^{+0.05}_{-0.04}$	&	$-0.6^{+1.1}_{-1.0}$	&	$(11.8^{+0.9}_{-11.6})$	&	$229.7^{+17.6}_{-226.5}$	&	48	&	HBC\,639	&	$1.7^{+0.1}_{-0.1}$	&	$>1.4$	&	$<1.9$	&	$<0.24$	\\
17	&	HD\,45677	&	$138.8^{+2.6}_{-2.7}$	&	$0.71^{+0.01}_{-0.01}$	&	$25.6^{+1.2}_{-1.0}$	&	$7.1^{+1.9}_{-1.9}$	&	49	&	Elias\,2-24	&	$2.0^{+0.2}_{-0.2}$	&	$0.60^{+0.02}_{-0.02}$	&	$5.7^{+0.7}_{-0.7}$	&	$0.68^{+0.09}_{-0.09}$	\\
18	&	VY\,Mon	&	$26.9^{+0.4}_{-0.4}$	&	$1.33^{+0.02}_{-0.02}$	&	$8.4^{+0.1}_{-0.1}$	&	$6.7^{+2.5}_{-2.5}$	&	50	&	Elias\,2-28	&	$3.4^{+0.5}_{-0.5}$	&	$0.44^{+0.02}_{-0.01}$	&	$9.7^{+2.1}_{-1.8}$	&	$1.2^{+0.3}_{-0.2}$	\\
19	&	HD\,259431	&	$6.30^{+0.1}_{-0.1}$	&	$1.48^{+0.08}_{-0.07}$	&	$4.2^{+0.1}_{-0.1}$	&	$2.8^{+0.4}_{-0.4}$	&	51	&	Elias\,2-30	&	$1.8^{+0.6}_{-0.5}$	&	$0.42^{+0.02}_{-0.02}$	&	$61.8^{+17.7}_{-16.6}$	&	$7.4^{+2.1}_{-2.0}$	\\
20	&	R\,Mon	&	$20.1^{+0.5}_{-0.4}$	&	$0.83^{+0.01}_{-0.01}$	&	$13.0^{+0.4}_{-0.4}$	&	$9.9^{+3.9}_{-3.9}$	&	52	&	V2129\,Oph	&	$1.7^{+0.2}_{-0.2}$	&	$0.51^{+0.02}_{-0.01}$	&	$9.1^{+1.4}_{-1.2}$	&	$1.1^{+0.2}_{-0.2}$	\\
21	&	HD\,50138	&	$57.4^{+1.0}_{-0.9}$	&	$0.97^{+0.01}_{-0.01}$	&	$19.4^{+0.3}_{-0.3}$	&	$7.6^{+1.7}_{-1.7}$	&	53	&	V2062\,Oph	&	$1.1^{+0.1}_{-0.1}$	&	$0.67^{+0.04}_{-0.03}$	&	$3.4^{+0.5}_{-0.5}$	&	$0.41^{+0.06}_{-0.07}$	\\
22	&	HD\,72106	&	$2.2^{+0.2}_{-0.2}$	&	$0.50^{+0.01}_{-0.01}$	&	$20.2^{+1.4}_{-1.4}$	&	$5.6^{+1.8}_{-1.8}$	&	54	&	HD\,150193	&	$13.7^{+0.3}_{-0.3}$	&	$0.67^{+0.01}_{-0.01}$	&	$15.9^{+0.3}_{-0.4}$	&	$3.4^{+1.1}_{-1.1}$	\\
23	&	HD\,87643	&	$226.3^{+9.5}_{-9.7}$	&	$0.72^{+0.01}_{-0.01}$	&	$37.3^{+1.3}_{-1.2}$	&	$56.0^{+18.8}_{-18.7}$	&	55	&	AK\,Sco	&	$3.3^{+0.2}_{-0.2}$	&	$0.63^{+0.03}_{-0.03}$	&	$8.0^{+1.4}_{-1.0}$	&	$0.8^{+0.2}_{-0.2}$	\\
24	&	CR\,Cha	&	$1.1^{+0.1}_{-0.1}$	&	$0.61^{+0.02}_{-0.02}$	&	$5.2^{+0.6}_{-0.6}$	&	$0.83^{+0.13}_{-0.13}$	&	56	&	KK\,Oph	&	$9.4^{+0.5}_{-0.5}$	&	$0.54^{+0.01}_{-0.01}$	&	$12.1^{+0.9}_{-1.0}$	&	$1.9^{+0.4}_{-0.4}$	\\
25	&	HD\,95881	&	$5.8^{+0.3}_{-0.2}$	&	$>1.3$	&	$<3.4$	&	$<0.6$	&	57	&	51\,Oph	&	$17.2^{+0.4}_{-0.5}$	&	$1.9^{+0.1}_{-0.1}$	&	$11.2^{+0.3}_{-0.2}$	&	$1.40^{+0.05}_{-0.05}$	\\
26	&	DI\,Cha	&	$1.8^{+0.1}_{-0.1}$	&	$0.55^{+0.01}_{-0.01}$	&	$14.1^{+1.1}_{-1.0}$	&	$2.3^{+0.3}_{-0.3}$	&	58	&	HD\,163296	&	$16.0^{+0.6}_{-0.6}$	&	$0.96^{+0.04}_{-0.03}$	&	$8.1^{+0.5}_{-0.4}$	&	$0.96^{+0.11}_{-0.10}$	\\
27	&	HD\,97048	&	$3.5^{+0.2}_{-0.2}$	&	$0.56^{+0.01}_{-0.01}$	&	$26.5^{+1.4}_{-1.5}$	&	$4.2^{+0.5}_{-0.5}$	&	59	&	HD\,169142	&	$0.7^{+0.2}_{-0.2}$	&	$0.41^{+0.02}_{-0.01}$	&	$83.0^{+19.3}_{-22.5}$	&	$12.0^{+3.1}_{-3.5}$	\\
28	&	HP\,Cha	&	$8.6^{+0.3}_{-0.3}$	&	$0.60^{+0.01}_{-0.01}$	&	$5.6^{+0.3}_{-0.3}$	&	$0.89^{+0.10}_{-0.10}$	&	60	&	MWC\,297	&	$125.2^{+2.1}_{-1.7}$	&	$>3.0$	&	$<46.3$	&	$<13.9$	\\
29	&	FM\,Cha	&	$2.7^{+0.3}_{-0.3}$	&	$0.52^{+0.01}_{-0.01}$	&	$13.5^{+1.6}_{-1.6}$	&	$2.2^{+0.3}_{-0.3}$	&	61	&	MWC\,300	&	$76.5^{+2.0}_{-1.9}$	&	$0.568^{+0.004}_{-0.004}$	&	$30.1^{+0.8}_{-0.8}$	&	$54.1^{+6.2}_{-6.2}$	\\
30	&	WW\,Cha	&	$5.2^{+0.3}_{-0.4}$	&	$0.67^{+0.42}_{-0.05}$	&	$5.4^{+1.5}_{-3.3}$	&	$0.9^{+0.3}_{-0.5}$	&	62	&	R\,CrA	&	$82.1^{+1.6}_{-1.6}$	&	$0.84^{+0.01}_{-0.01}$	&	$14.1^{+0.3}_{-0.3}$	&	$1.8^{+0.3}_{-0.3}$	\\
31	&	CV\,Cha	&	$2.2^{+0.1}_{-0.1}$	&	$0.61^{+0.02}_{-0.02}$	&	$6.1^{+0.6}_{-0.6}$	&	$0.98^{+0.14}_{-0.13}$	&	63	&	T\,CrA	&	$5.9^{+0.6}_{-0.6}$	&	$0.49^{+0.02}_{-0.02}$	&	$19.8^{+5.6}_{-4.2}$	&	$2.6^{+0.8}_{-0.7}$	\\
32	&	HD\,98922	&	$24.8^{+0.6}_{-0.6}$	&	$1.12^{+0.03}_{-0.02}$	&	$8.1^{+0.2}_{-0.2}$	&	$9.3^{+4.2}_{-4.2}$	&	64	&	HD\,179218	&	$20.1^{+1.5}_{-1.3}$	&	$0.55^{+0.01}_{-0.01}$	&	$27.7^{+1.9}_{-2.0}$	&	$7.0^{+1.2}_{-1.2}$	\\

\hline
 
 \end{tabular}
\tablefoot{Numbers between parentheses need to be multiplied by $10^3$.}
}
\end{table*}

Other fits of objects indicate intrinsic discrepancies from the proposed disk geometry. Notable examples are 51\,Oph and MWC\,297. A closer inspection of the fits shows that the departure is at least partly related to how close the inner rim can get to the central star. In our disk model, it is assumed that no emission comes from $R<R_\mathrm{sub}$, where $R_\mathrm{sub}$ is calculated following Eq.~(\ref{eq:rsub}). The assumption puts a limit on how compact the disk emission can be, which translates into an upper limit on the correlated-flux profile. It is clear that the two mentioned objects are fit by a model with too many visibility nulls, which physically corresponds to model disks that have an inner radius that is too extended to reproduce the compact emission. Interestingly, this suggests that the considered objects have a strong mid-infrared contribution from within the dust sublimation radius (see, e.g., \citealt{2008A&A...485..209A}).

Finally, there are three objects with a mid-infrared contribution that is almost entirely stellar: $\beta$\,Pic, HD\,109085, and V2246\,Oph. As is clear from the model plots, all but the 0-baseline fluxes are consistent with the predicted photospherical fluxes. In other words, any mid-infrared excess is only detected in the total flux observations, and the excess is resolved out in the interferometric observations. The first two objects are well-known debris-disk objects (e.g, \citealt{1984Sci...226.1421S,2005ApJ...620..492W}), and the MIDI data give lower bounds on the dust location (see \citealt{2009A&A...503..265S}, for HD\,109085). V2246\,Oph, on the other hand, may even lack a disk, as pointed out by \citet{2009ApJ...703..252J}. The fits of these three objects are poorly constrained.

\section{Mid-infrared size-luminosity relation}\label{sect:sizelum}
The temperature-gradient model in Sect.~\ref{sect:results} is expressed in terms of two free parameters: the total disk flux $F_{\mathrm{disk,}\nu}$ and the temperature gradient $q$. The physical relevance of the first parameter is clear, but the second parameter needs some more attention.

A typical temperature profile $T\propto R^{-1/2}$ is found in classical models for the structure of flaring protoplanetary disks (e.g., \citealt{1987ApJ...323..714K}). A different, limiting geometry for disks is a geometrically thin, passive irradiated disk, where a temperature profile $T\propto R^{-3/4}$ can be derived, see, for example, \citet{2007astro.ph..1485A}. As mentioned by the latter author, this is probably the steepest profile one could expect from a passive disk. 

The temperature gradients we derive span a much wider range. It is therefore possibly incorrect to interpret the parameter as a real temperature gradient within the disk. On the one hand, the model description is too simplistic to capture the actual intensity distribution of the disk. On the other hand, the temperature gradient is the only model parameter for (implicitly) describing the geometry, so it may capture much more than the dust temperature alone.

\begin{figure*}
 \centering
 \includegraphics[width=.9\textwidth,viewport=50 0 930 400,clip]{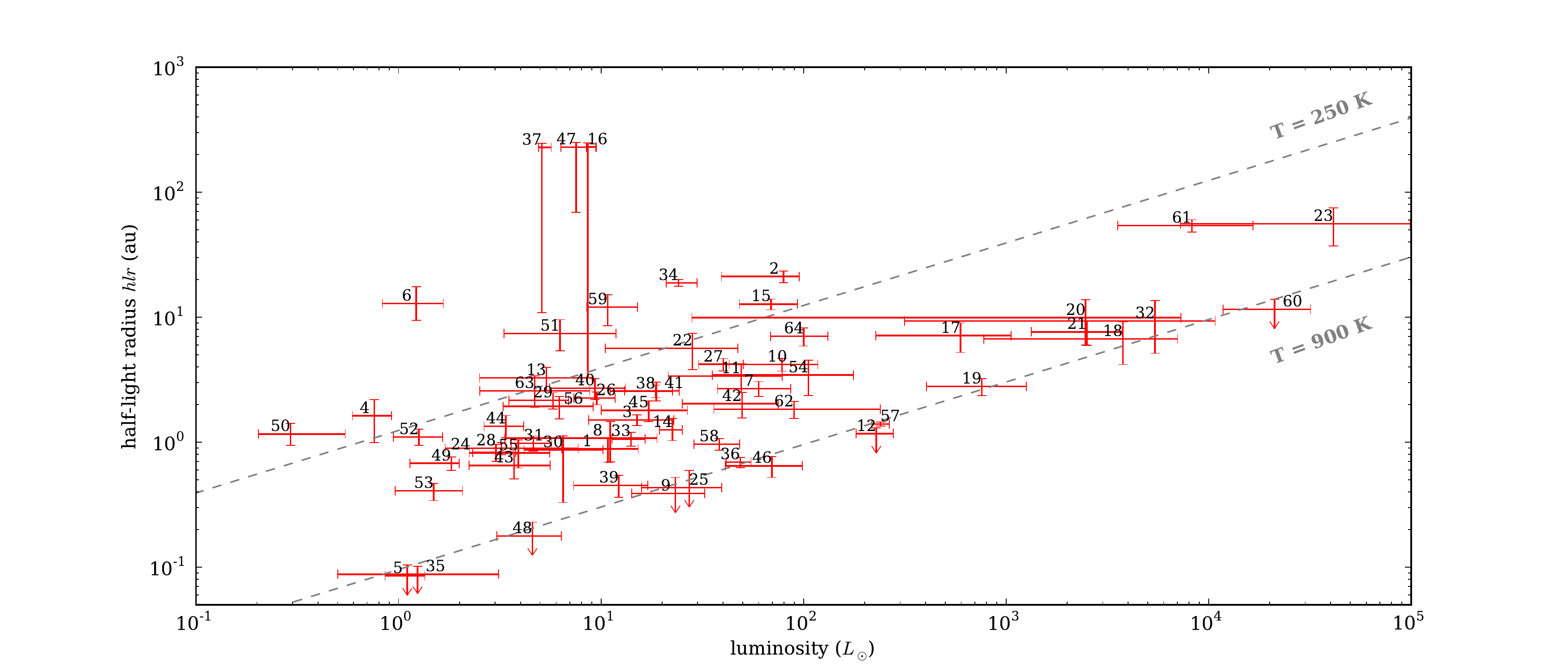}
 \caption{MIDI size-luminosity diagram for protoplanetary disks: plot of the half-light radii of the disks at $10.7\,\mu$m vs.~the stellar luminosity of the sample targets. The gray dashed lines indicate the expected distance at which gray, optically thin dust at the indicated temperature would be located (cf.~\citealt{2009ApJ...700..491M}). }\label{fig:sizelum}
\end{figure*}

An alternative interpretation would be to see $q$ acting as a radial brightness parameter. Under this point of view, we define a half-light radius $hlr$, as follows:
\begin{equation}
 \frac{F_{\mathrm{disk},\nu}}{2}=\int_{R_\mathrm{sub}}^{hlr}\mathrm dR\,2\pi R\,\,I_{\mathrm{disk},\nu}(R).
\end{equation}
In words, $hlr$ is the disk radius within which half of the mid-infrared flux is emitted. Since disks may have gaps, which result in a discontinuity in the radial brightness profile, the use of a half-light radius has a clearer interpretation than a parameter that imposes continuity.

The interest in a size parameter, like $hlr$, is that we can compare it directly with stellar parameters. In near-infrared interferometry, it is common to compare disk sizes with the stellar luminosity (see \citealt{2007prpl.conf..539M}, and references therein). The size-luminosity relation we derive from our mid-infrared interferometric data is shown in Fig.~\ref{fig:sizelum}. Across a luminosity range of almost seven orders of magnitude, we see a gradual increase in the size of the targets, with considerable scatter.\footnote{We should note that the term ``size'' only points to an apparent size: intrinsically, the models all have the same outer radius ($300\,$au, see Sect.~\ref{sect:model}).}

The majority of the objects lie within the range defined by the expected radii at which gray, optically thin dust between 900\,K and 250\,K would be located. A handful of objects seems to be ``oversized'' with respect to this broad range. Three peculiar objects in the plot are HD\,109085 (\#37), V2246\,Oph (\#47), and $\beta$ Pic (\#16), which have very poor constraints on their size. As pointed out in Sect.~\ref{sect:results}, these objects only show a mid-infrared excess in the total flux, and the exact ``location'' of their disks is poorly constrained: based on the data, only a lower limit on the half-light radius can be derived.\footnote{The upper limit on $hlr$ is put by the assumed model, which has an outer radius of $300\,$au.}  A priori, it could also have been expected that these debris-disk (or maybe even diskless, for V2246\,Oph) objects are atypical. Interestingly, several other oversized objects are well known (pre-)transitional disks (e.g., HD\,169142 (\#59), \citealt{2012ApJ...752..143H}; HD\,100546 (\#34), \citealt{2010A&A...511A..75B}; GM\,Aur (\#6), \citealt{2009ApJ...698..131H}), which supports the actual relevance of our size parameter.

\citet{2009ApJ...700..491M} studied a sample of, on average, more massive protoplanetary disk objects using mid-infrared sparse-aperture interferometry on a 10 m class telescope. Although a different model geometry was used (a ring model), the representative size range they report is very similar (our choice to show the 900-K and 250-K lines follows their work). Since typical visibilities of the single-telescope observations are above 0.7 (or, $V^2>0.5$), the specific model plays a minor role, and a direct comparison of the size ranges is justified. We thus conclude that, on both large scales (probed by the sparse-aperture interferometry) and small scales (probed by the MIDI data), the sources look similar. 

As already pointed out by \citet{2009ApJ...700..491M}, the rather poor correlation between stellar luminosity and mid-infrared disk size strongly contrasts with the tight correlation found between luminosity and near-infrared disk sizes \citep{2002ApJ...579..694M,2005ApJ...624..832M,2007prpl.conf..539M}. In the near-infrared, the radiation is dominated by dust near the sublimation temperature, located at the inner rim of the disk. For a given sublimation temperature, the location of this sublimation radius is (to first order) entirely determined by the stellar luminosity (Eq.~(\ref{eq:rsub})), from which the tight correlation immediately follows. The region responsible for the mid-infrared emission, on the other hand, is much more extended. The scatter in Fig.~\ref{fig:sizelum} then indicates a large variety in disk geometries. \citet{2009ApJ...700..491M} mention a few physical differences that may explain the different appearance of disks around similar central stars: radial variations in grain distribution, grain growth and settling, and binarity. We already mentioned above that several of the apparently over-sized objects are transition disks. In Sect.~\ref{sect:sizecolor}, we investigate the origin of the different disk sizes.

In conclusion, our mid-infrared size-luminosity relation extends the existing size-luminosity relations by an order of magnitude down in minimal size, toward the low-mass star formation domain (for the massive star regime, see \citealt{2011A&A...532A.109G} and \citealt{2013A&A...558A..24B}).

\begin{table}
\caption{Classification of the Herbig Ae stars in the sample, following the \citet{2001A&A...365..476M} classification. }\label{table:meeus}
\begin{tabular}{ll}
\hline\hline
group&sources\\
\hline
Ia&V892\,Tau, AB\,Aur, HD\,36112, CQ\,Tau, HD\,38120,\\& HD\,259431,  HD\,100546, HD\,139614, HD\,142527, \\&R\,CrA, T\,CrA, HD\,179218\vspace{.2cm}\\
Ib&V1247\,Ori, HD\,97048,  HD\,100453, HD\,135344\,B,\\& Elias\,2-30, HD\,169142\vspace{.2cm}\\

IIa&RY\,Tau, SU\,Aur, HD\,31648, UX\,Ori, HD\,36917\tablefoottext{$\ast$}, \\& HD\,72106, HD\,95881, HD\,98922, HD\,104237,\\& HD\,142666, HD\,144432, HD\,144668, HD\,150193,\\& AK\,Sco, KK\,Oph, 51\,Oph, HD\,163296\\

\hline

\end{tabular}\\
\tablefoottext{$\ast$}{HD\,36917 may lack any silicate feature and be a group IIb source, according to \citet{2010ApJ...721..431J}. Our MIDI spectrum for this target points to a weak silicate feature.}

\end{table}

\section{Size-color relations for Herbig Ae disks}\label{sect:sizecolor}
Pre-main sequence stars with similar luminosities turn out to have disks that are different. If all objects have the same age, the difference is a manifestation of the intrinsic diversity of the disks (e.g., in mass, viscosity, angular momentum). Conversely, if the objects have different ages, the diversity may hint at different evolutionary stages. The diversity in disk sizes probably points to a combination of the two and forms the topic of this section.

\subsection{Herbig Ae disks}
More than half of the objects in our sample are categorized as Herbig Ae stars, and within our sample this may be the only group of objects with enough representatives for relevant comparisons. We limit our further analysis to this group of objects. The interest in this group arises from the quoted evolutionary scenarios presented in Sect.~\ref{sect:intro} and Fig.~\ref{fig:scenarios}.\footnote{Although the original works propose their scenarios as representative for Herbig Ae/Be stars, their actual source lists are mostly limited to Herbig Ae objects.} 

The Herbig Ae stars can be classified following the \citet{2001A&A...365..476M} scheme: I = the ``flaring'' disks, and II = the ``flat'' disks. An index ``a'' and ``b'' is used to indicate whether a 10-$\mu$m silicate feature is seen or not. The classification for the Herbig Ae stars in the sample is shown in Table \ref{table:meeus}.

\begin{table}
\caption{Parameters of the radiative-transfer model population.}\label{table:parameters}
 \begin{tabular}{ll}
  \hline\hline
  parameter&value\\\hline
  \textbf{star}\\
  mass/lum./eff.~temp.  & (1.7\,$M_\odot$, 11\,$L_\odot$, 7000\,K),\\
  \,\,($M_\star$, $L_\star$,$T_\mathrm{eff}$)&(2.0\,$M_\odot$, 21\,$L_\odot$, 8500\,K),\\
  &(2.6\,$M_\odot$, 60\,$L_\odot$, 10000\,K),\\
  &(3.5\,$M_\odot$, 200\,$L_\odot$, 13000\,K)\\
  distance $d$&140\,pc\\\\
  \textbf{disk}\\
  surface density power $p$&1.0\\
  dust mass $M_\mathrm{dust}$&$10^{-3.5}\,M_\odot$\\
  disk inner radius $R_\mathrm{in}$& $R_\mathrm{sub}$, 1, 2.5, 4, 6, 8,  10, 15, 20\,au\\
  disk outer radius $R_\mathrm{out}$& 300\,au\\
  minimal grain size $a_\mathrm{min}$&0.01, 0.03, 0.1, 0.3, 1.0, 3.0\,$\mu$m\\
  maximal grain size $a_\mathrm{max}$&1\,mm\\
  grain size distribution & -3.5\\
  dust composition&amorphous MgFeSiO$_4$ (90\,\%)\\&carbonaceous grains (10\,\%)\\
  dust-to-gas ratio&0.01\\
  turbulence parameter $\alpha$&$10^{-5},10^{-4},10^{-3},10^{-2}$\\
  inclination $i$&10, 25, 40, 60, 70$^\circ$\\\\
  \textbf{halo}\\
  halo inner radius&$R_\mathrm{sub}$\\
  halo outer radius&$1.3\,R_\mathrm{sub}$\\
  dust composition halo&same as disk, $a=0.3-1.0\,\mu$m\\
  halo opacity $\tau_V$&0.2\\\\
  \multicolumn{2}{l}{\textbf{interferometric model observations}}\\
  baseline lengths&10, 20, 30, \ldots, 130, 140\,m\\
  baseline angles&0, 15, 30, \ldots, 150, 165$^\circ$\\\hline
  
 \end{tabular}

\end{table}

\subsection{Size-color diagram}
\begin{figure}
 \centering
 \includegraphics[width=.5\textwidth,viewport=20 10 530 400,clip]{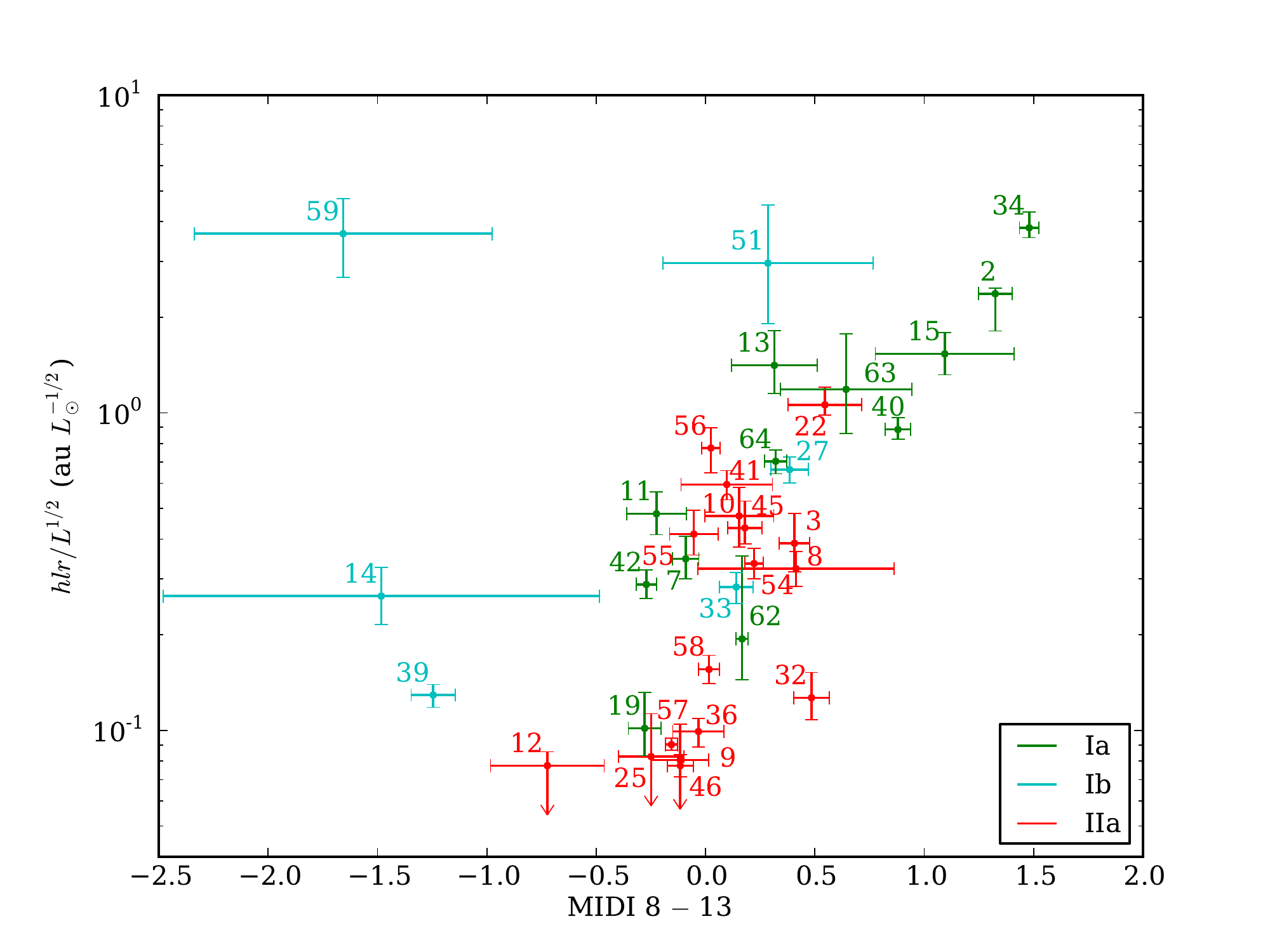}
 \caption{Size-color diagram for the Herbig Ae stars within the sample: plot of the half-light radius ($hlr$) of the disks, normalized to the luminosity $L^{1/2}$, versus the $N$-band continuum slope (calculated as $-2.5\,\log(F_{\nu,8\,\mu\mathrm m}/F_{\nu,13\,\mu\mathrm m}$). The colors of the points and errorbars indicate the \citet{2001A&A...365..476M} classification of the sources. }\label{fig:sizecolor}
\end{figure}

In Fig.~\ref{fig:sizecolor} we compare the half-light radius to the MIDI $8-13\,$color, calculated as $-2.5\,\log(F_{\nu,8\,\mu\mathrm m}/F_{\nu,13\,\mu\mathrm m})$. This color is an estimate of the continuum spectral slope in the $N$ band. To compare the objects, which have different luminosities, we normalize the half-light radii by $L^{1/2}$, the expected scaling between characteristic sizes and luminosities (Fig.~\ref{fig:sizelum}). We note that the quantities in the plot are distance-independent.\footnote{Both the estimates for $hlr$ and $L^{1/2}$ scale linear with the distance, and colors are obviously independent of distance.} As is clear, sources with a blue $N$-band color tend to be small, while large objects have a red color, on average. Adding a color to the size comparison thus seems to partly decouple the degeneracy between sizes and luminosities.

\citet{2004A&A...423..537L} derive the mid-infrared sizes of seven Herbig Ae disks using MIDI interferometry and notice a similar connection with the mid-infrared disk color. These authors interpret the correlation as evidence for the distinction of group I and group II by flaring versus flat disks: flaring disks will appear larger, and the colder outer emitting disk surface will make the disk redder. The interpretation follows the classical idea of \citet{2001A&A...365..476M}. 

In the size-luminosity diagram (Fig.~\ref{fig:sizelum}), it is made clear that some objects are oversized. We pointed out that several of these sources are confirmed to have a gap (Sect.~\ref{sect:sizelum}). The presence of radial gaps in the disk structure is not included in the flared- versus flat-disk picture and can be expected to alter the size-color relation.

\subsection{Radiative-transfer models}\label{sect:radtran}
To disentangle the multiple effects that play a role in our size-color diagram, we simulated a population of disk models with various parameters. We used the Monte Carlo radiative-transfer code \texttt{MCMax} \citep{2009A&A...497..155M} for this purpose. The code solves the temperature structure and vertical hydrostatic equilibrium in an azimuthally symmetric disk. Additionally, a grain-settling mechanism based on the strength of turbulence is implemented, parametrized by the $\alpha$-parameter (for classical $\alpha$ disks). Lowering $\alpha$ decouples the larger grains and settles them to the midplane.

The assumed model is a protoplanetary disk with a varying inner radius and a fixed outer radius ($R_\mathrm{out}=300\,$au, i.e., like the temperature gradient in Sect.~\ref{sect:model}), consisting of amorphous silicates of olivine stoichiometry and carbonaceous grains (\citealt{1995AA...300..503D,1993A&A...279..577P}, respectively). Of the potentially high number of parameters that can be varied, we limited ourselves to varying the parameters that have a strong influence on simulating the $N$-band interferometry of disks. The other parameters were fixed to representative values for Herbig Ae stars (e.g., \citealt{2012A&A...539A...9M}). Standard radiative-transfer models typically have problems reproducing the strong near-infrared excess of Herbig Ae stars. \citet{2012A&A...539A...9M} discuss different methods for adapting standard disk models in order to reproduce this near-infrared bump and propose the use of a compact halo component in addition to the disk. As these authors note, the compact halo is a parametrization of emission from the inner regions, including dust or gas within the inner rim. Although the near-infrared excess is not modeled here, the contribution of this component extends to the $N$ band. We include a similar halo component here, whose properties are fixed. For models without a gap, the inner part of the disk overlaps with the halo, and for models with a gap, the gap is represented by the (dust-free) region in between the halo and the disk.

An overview of the parameter grid is given in Table \ref{table:parameters}. For the model observations, we calculated a densely sampled UV plane (i.e., combination of baseline lengths and angles).

\subsection{Model half-light radii: inclination and baseline effects}\label{sect:incbase}
To place the radiative-transfer disk models into Fig.~\ref{fig:sizecolor} and compare them to the data, we simulated MIDI observations of the radiative-transfer models and then fit the same temperature-gradient disk model (Sect.~\ref{sect:model}) as we applied it to the data. The model interferometry thus allows us to derive half-light radii for the radiative-transfer models. Assuming that the parameter space for the radiative-transfer models is representative for the observed objects, two additional aspects in obtaining valid model comparisons are important: disk inclinations and baseline configurations.

\paragraph{Inclination.} The temperature-gradient model does not include inclination effects, i.e., the model is applied under the assumption that the observed disks (Sect.~\ref{sect:results}) and radiative-transfer models (this section) have a pole-on orientation. In reality, protoplanetary disks obviously have some inclination (though the Herbig Ae stars in our sample will not have a high inclination, since they are selected to have a low optical extinction). However, because the assumption of a pole-on orientation is applied to both the simulated observations of the radiative-transfer models and to the actual observations of the real disks, this does not introduce a bias in the comparison of the half-light radii. Furthermore, the half-light radius determined using a pole-on approximation is a robust parameter, and does not significantly depend on the intrinsic disk orientation (see Appendix \ref{appendix}).

\paragraph{Baseline configuration.} A related problem is the choice of the ``model baseline configuration''. The radial intensity profile of the temperature-gradient model will generally be different from that of the actual disk. This leads to half-light radii that depend on the specific baseline configuration (UV coverage). For example, the size estimate of an inclined disk along the major and minor axes differs. For each radiative-transfer disk model there is therefore a corresponding range of half-light radii.

To calculate this range of half-light radii for a given radiative-transfer model, we fit the temperature-gradient model to each individual UV point. Given that the full UV coverage of a single model contains 168 points (14 baseline lengths $\times$ 12 baseline angles), we thus get the same number of half-light radii, per model. The plots in the next section will show the range of the half-light radii we get.

The set of MIDI observations defines a baseline configuration for each target. Except for four objects, all Herbig Ae stars have at least two UV points. Having a UV coverage with several baseline lengths and baseline angles will lead to a more constrained value for the half-light radius, which averages over the size estimates of observations on a single baseline. Even if the temperature-gradient description is only approximative and assumes a pole-on orientation, our calculation of the model half-light radii, a parameter that is robust to inclination effects, represents a very conservative basis for comparing with the observed half-light radii.

\begin{figure*}
 \centering
 \includegraphics[width=.49\textwidth,viewport=10 10 530 410,clip]{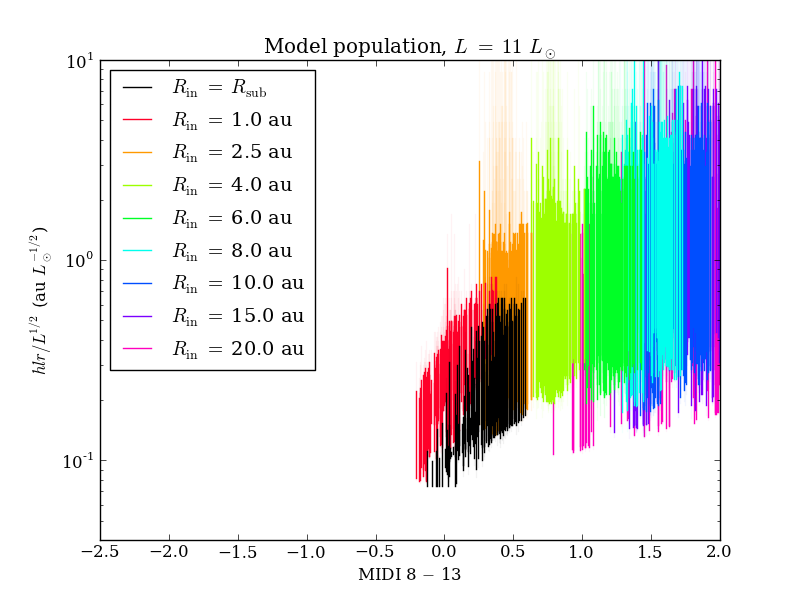}
 \includegraphics[width=.49\textwidth,viewport=10 10 530 410,clip]{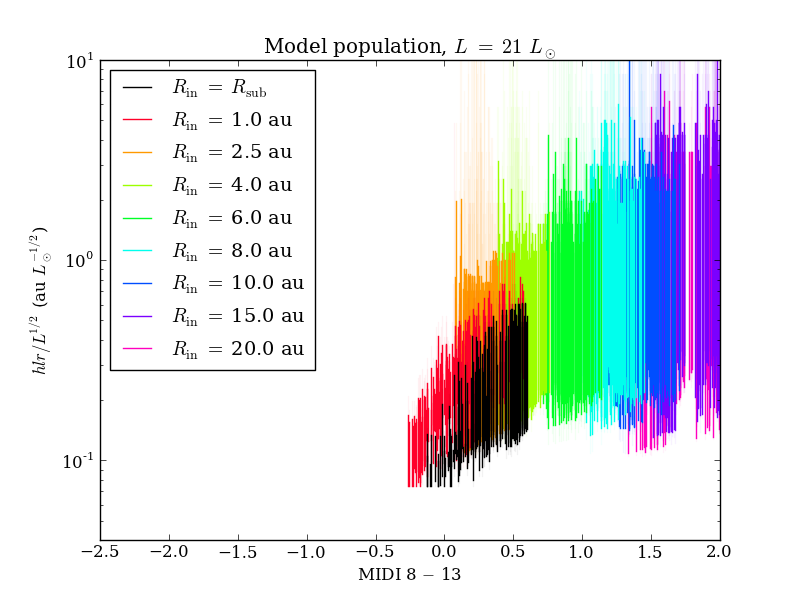}\\
 \includegraphics[width=.49\textwidth,viewport=10 10 530 410,clip]{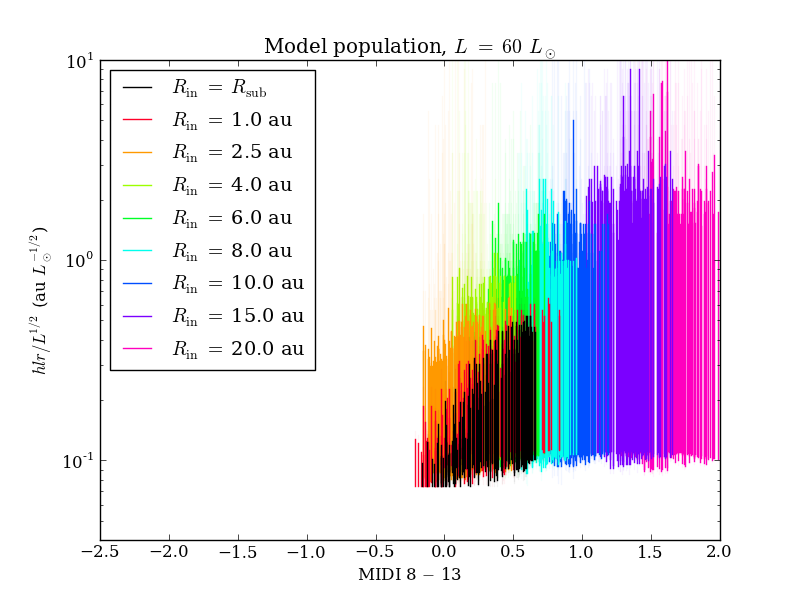}
 \includegraphics[width=.49\textwidth,viewport=10 10 530 410,clip]{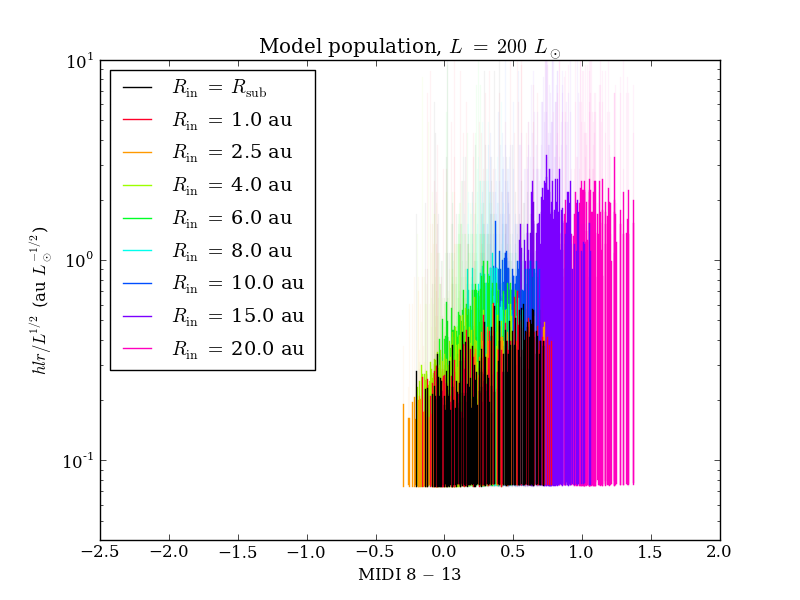}
 \caption{Size-color diagram for the population of radiative-transfer models for the four different luminosity cases ($11\,L_\odot$, $21\,L_\odot$, $60\,L_\odot$, and $200\,L_\odot$). The vertical lines indicate the range of sizes that can be obtained for the individual models. The 5\,$\%$ smallest and $5\,\%$ largest mid-infrared sizes are shown in transparent colors, giving an idea of the typical (90\,$\%$) range of model sizes. In black, the models without a gap are shown. The other colors show models with a gap between the halo and the disk components (for an increasing disk inner radius $R_\mathrm{in}$).}\label{fig:modelpop}
\end{figure*}

\subsection{Theoretical versus observed size-color relations}\label{sect:comparison}
Figure~\ref{fig:modelpop} shows the size-color diagram for the model population. The figure is split up into panels that show the four different stellar luminosities for which we have made radiative-transfer models. For all models, the range of half-light radii (see Sect.~\ref{sect:incbase}) is indicated. The axes ranges are identical to those in Fig.~\ref{fig:sizecolor}.

For all luminosity cases, the colors and sizes of models without a gap (i.e., with $R_\mathrm{in}=R_\mathrm{sub}$) spans a similar range (a broadening downward is seen for the highest luminosity case). This range is relatively compact: gapless disks can only have a limited range of half-light radii and mid-infrared colors. Introducing a gap in the disk by moving out the inner radius of the disk component substantially expands the possible range of colors and half-light radii. This occurs in a qualitatively similar fashion for all luminosities. As can be expected, large gaps will move the half-light radius outward, and the larger emitting surface of relatively low-temperature material makes the disk redder. Similar gap sizes correspond to bluer disk colors for more luminous central objects. This is a simple consequence of the disk wall being warmer for the more luminous central objects.

\begin{figure}
 \centering
 \includegraphics[width=.5\textwidth,viewport=20 10 530 400,clip]{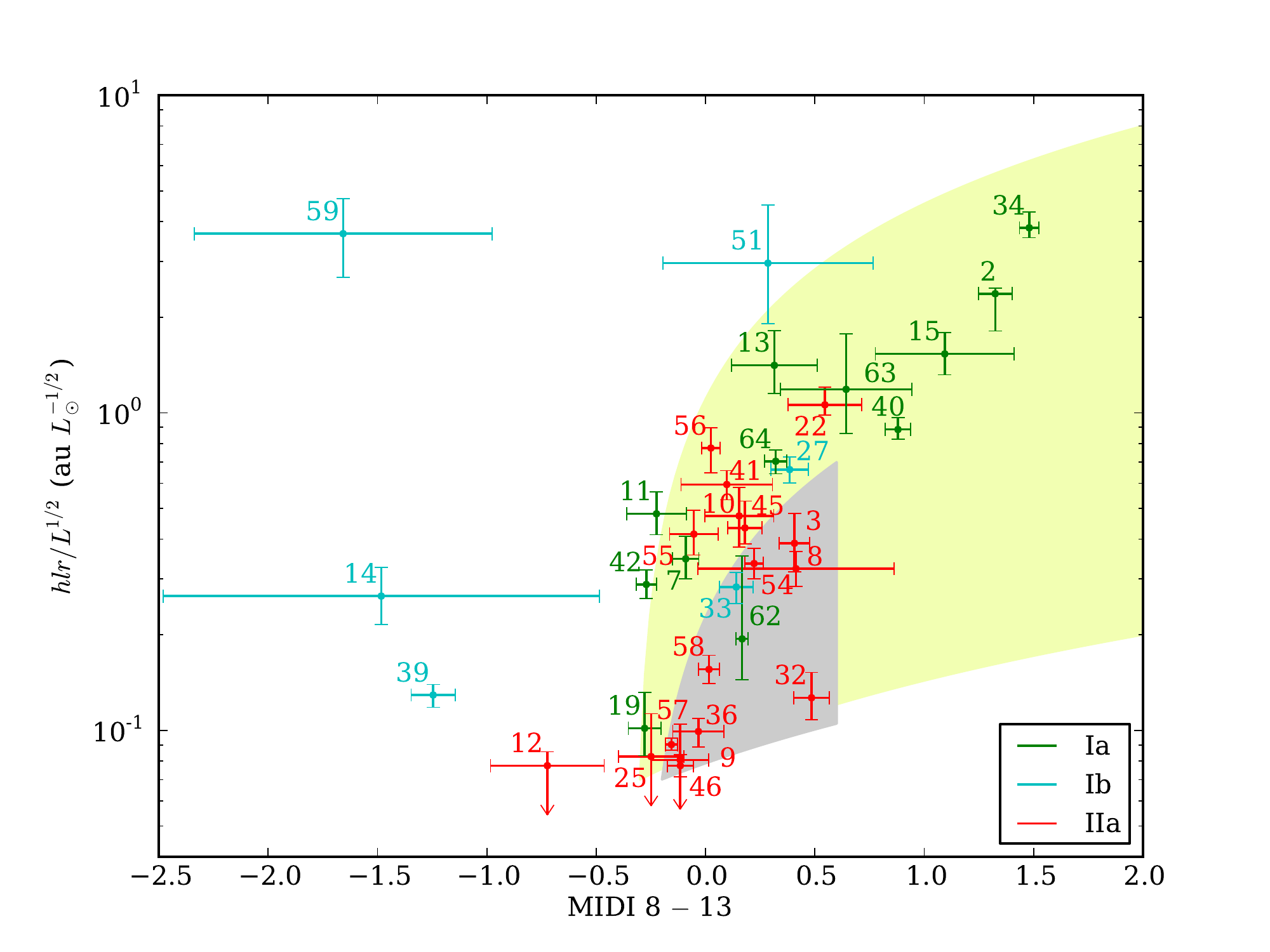}
 \caption{Merged size-color diagram, combining observations (Fig.~\ref{fig:sizecolor}) with the model population (Fig.~\ref{fig:modelpop}). In gray, we indicate the range of models without a gap. In yellow, we show the approximate range of models with a gap between the inner halo and the disk component. We note that both regions indeed overlap. Group Ib sources with colors deferring from the model population are V1247\,Ori (\#14), HD\,135344\,B (\#39), Elias 2-30 (\#51), and HD\,169142 (\#59).}\label{fig:mergedsizecolor}
\end{figure}

The model size-color plot allows us now to interpret the MIDI size-color diagram for Herbig Ae stars. In Fig.~\ref{fig:mergedsizecolor}, we combined the observational diagram (Fig.~\ref{fig:sizecolor}) with the approximate model ranges in Fig.~\ref{fig:modelpop}. Except for a few outliers, which are all in group Ib, the range defined by the observations largely corresponds to the model range. Focusing first on the models with a continuous disk (gray zone in Fig.~\ref{fig:mergedsizecolor}), we notice that only a limited number of sample stars have the size-color combination falling into this area of the diagram. The majority of these targets are group II sources. Of the group Ia group, most objects have characteristics corresponding to the gapped sources. They share this property with a significant number of the group II sources. 

\citet{2013A&A...555A..64M} suggest all group I sources may have gaps, whereas it is proposed that group II sources have continuous, gapless disks. Our size-color diagram not only confirms that most group I objects have (inner) disks that differ from standard, continuous disks. It also shows in a simple way that there may be a significant overlap in structure of the inner disk region for both groups or, at least, for group Ia and group II objects. A comparison of the overlap region with the model diagrams even suggests that the structure of these group II disks differs from that of continuous disks, \emph{and they may have gaps}. Gaps in group II disks would alter the picture that group I and group II disks form two strictly separated populations. The above findings are discussed further in Sect.~\ref{sect:discussion}.

\subsection{Non-thermal mid-infrared emission}\label{sect:nonthermal}
Objects that clearly are not represented by our model population are the group Ib disks with very blue colors. Group Ib objects lack the $N$-band silicate feature by definition, a property that is associated with their large gaps in the radial dust distribution (e.g., \citealt{2012ApJ...752..143H,2013A&A...555A..64M,2014A&A...567A..51C}). Instead of the silicate feature, their $N$-band spectra typically have strong emission peaks of polycyclic aromatic hydrocarbons (PAHs; \citealt{2010ApJ...721..431J}). These peaks are intimately related to the gap nature: the decrease in dust emission makes it easier to observe the PAH emission \citep{2013A&A...555A..64M}, and the ionization imprint in the PAH emission traces the density throughout the disk \citep{2014A&A...563A..78M}. 

The emission mechanism of PAHs is based on the electronic excitation by ultraviolet photons and the subsequent emission in molecular stretching and bending modes. In (and just outside) the $N$ band, emission peaks of these modes are centered at 7.7, 8.6, 11.3, and 12.7\,$\mu$m. The transient heating mechanism, a property that PAHs share with very small grains (VSGs), is a potential source of mid-infrared emission coming from the outer disk regions. 

The effect of PAHs and VSGs on $N$-band interferometry is twofold. First, a strong PAH contribution, notably around the typically strong and wide 8\,$\mu$m band, makes the MIDI $8-13$ color bluer. Second, a strong mid-infrared contribution of PAHs or VSGs in the outer disk will make a disk appear much larger than if its emission were thermal. The relative contribution of these molecules or grains to the $N$ band is the strongest for group Ib objects, which explains why they have deviant disk colors and sizes in Fig.~\ref{fig:sizecolor}. In a sense, the connection between apparent disk size and color, usually coupled by the grain temperature of particles in thermal equilibrium with the radiation field, is decoupled by the stochastically heated grains.

\subsection{Limitations of the model population}\label{sect:limitations}
A direct comparison of the diagram suggests that the model population covers the necessary range of parameters to interpret our observations. However, we need to be careful with the interpretation. 

The model population consists of two-component models with a halo and a disk, which can be decoupled by a gap. This model is representative for some Herbig Ae disks, but certainly not for all. Notable exceptions are the transitional disks that still have a massive inner disk. Example objects are AB\,Aur (Ia, \citealt{2010ApJ...718L.199H}) and HD\,97048 (Ib, \citealt{2013A&A...555A..64M}). Including these more complex disk structures is, at least in principle, possible in the disk population, but introduces several new parameters. The latter would increase the number of models, hence the model computation time, significantly. The extra models also will not give any new insight: the observational diagram is already fully covered by the models (apart from some group Ib objects), which indicates that new models will only point to model degeneracies.

This point brings us to the main limitation of our comparison: mid-infrared sizes and colors are (obviously) insufficient for assessing the full disk structure. The following problems can be identified:
\begin{itemize}
 \item A fixed dust composition: our models use a fixed dust composition. Varying the silicate composition and/or the continuum opacities may slightly change the MIDI $8-13$ color.
 \item A fixed halo component: the halo component is kept fixed for the model population. Making the halo component weaker will make the near-infrared excess disappear, which is excluded by typical Herbig Ae SEDs \citep{2012A&A...539A...9M}. Increasing this component makes the mid-infrared emission bluer and more compact.
 \item Gaps in the outer disk: the mid-infrared wavelength range is only sufficient for exploring gaps in the inner disks. Gaps in the outer disk will only be visible at longer wavelengths. In the absence of any other gaps, a disk will appear as standard in the $N$ band.
 \item Gap sizes: the gap size in a given part of the model diagram depends on the luminosity of the central star. When translating the color and size of an object into a possible gap size, one needs to take the luminosity of the central star into account.
 \item Model size range: the range of sizes (or half-light radii) that can correspond to a given color seems much larger from the gapped-disk models than what is actually observed. This is mainly a model bias. As explained in Sect.~\ref{sect:incbase}, each disk model has a maximum range of model sizes. Having a few observations, which is typically the case for our MIDI data, strongly confines the size estimate of a disk (Sect.~\ref{sect:incbase}). 
\end{itemize}
In summary, our model population might partly underestimate the range in colors that disks can have, and it also has the tendency to overestimate the typical size range of disks. 

These limitations of the model population may hamper a direct comparison with the MIDI size-color diagram. However, it is clear that most effects are not changing our main conclusions in Sects.~\ref{sect:comparison} and \ref{sect:nonthermal}: some group II disks are incompatible with standard disks, and the contribution of stochastically heated grains is needed to interpret the diagram.

\section{Tracing the evolution of Herbig Ae disks}\label{sect:discussion}
Ideally, understanding the structure of Herbig Ae disks in the mid-infrared demands a detailed individual modeling of a large sample of targets with a high number of observations at high spatial resolution. A large number of objects have indeed been observed with mid-infrared interferometry, but the number of observations per target is generally too low for constraining sophisticated models. We modeled the existing, archival observations in a simple way and interpret the results in this section.

\subsection{Overlap in spatial properties group I and II}
Protoplanetary disks in hydrostatic equilibrium that are irradiated by a central star will have a limited range of possible mid-infrared colors and (apparent) sizes, determined by the reprocessing of the stellar radiation by the dust. Under the reasonable assumption that the objects are largely azimuthally symmetric, any deviation from the standard color-size relation must point to differences in the radial and/or vertical structure of the disk. 

The idea that most group I objects may possess large radial gaps is almost exclusively based on observational evidence from the outer disks around Herbig Ae stars \citep{2012ApJ...752..143H,2013A&A...555A..64M}. Our mid-infrared interferometric analysis of disks provides more insight into the nature of the inner disks of group I objects. From our comparison between the observed sizes and colors of Herbig Ae disks and sizes and colors of radiative-transfer models in Sect.~\ref{sect:sizecolor}, we found that most group I objects cannot be interpreted as having ``standard'' continuous disks. They fall in a region where the radiative-transfer models have gaps in their radial dust distribution. This direct evidence for gaps in the inner disks of group I objects endorses the idea that most of these disks are transitional. Whether all group I disks in our sample have gaps cannot be confirmed, since the observations of at least a few objects (e.g., R\,Cra (\#62)) are in agreement with there being a continuous inner disk.

For the group II objects, Fig.~\ref{fig:mergedsizecolor} indicated that several have colors and sizes that agree with models for continuous disks. This agrees with the classical picture that group II disks are gapless and flat. However, a number of group II sources differ from ``standard'' continuous disks. They fall in an overlapping region with the gapped group I sources. Also the radiative-transfer models in the overlap regions possess gaps in their radial dust distribution. Is it correct to interpret these group II sources as having a gap? For at least two group II disk sources, gaps have already been claimed. Based on near- and mid-infrared interferometric observations (IOTA and MIDI, respectively) \citet{2013A&A...555A.103S} modeled the disk around HD\,142666 as having a gap with a radius of $0.8\,$au, plus an additional inner disk between $0.3$ and $0.35\,$au. \citet{2012A&A...541A.104C}, on the other hand, observed and analyzed the disk around HD\,144432, and their interferometric observations (AMBER and MIDI) point to a similar model with a gap radius of 1\,au and a compact 0.2-au radius inner ring. The two sources are group II sources in the overlap region between group I and II sources (\#41 and \#45), and both sources fall outside of the range of disk models without gaps.

In addition to the above examples, millimeter observations of the group II source RY\,Tau indicate that the inner disk environment may be dust-depleted \citep{2010ApJ...714.1746I}. The lack of millimeter emission for this object may not immediately imply a physically gapped disk (e.g., \citealt{2012ApJ...750..161D}). Still, in the size-color diagram in Fig.~\ref{fig:mergedsizecolor}, RY\,Tau (\#3) lies at the top of the region that is compatible with disk models without gaps. This object may therefore be an example of a gapped group II disk that is missed by the simple size and color criteria. Our conservative way of estimating the possible range of half-light radii may in fact impede the identification of RY\,Tau as a gapped source, and even more group II objects with large half-light radii may have gaps.

We conclude that, although the direct translation of sizes and colors in the 
presence of gaps in group II disks might not be straightforward, there seems to 
be a promising correspondence between the two. The discovery that also 
some group II disks have gaps alters the existing picture of group I and II disks.

\begin{figure}
 \centering
 \includegraphics[width=.5\textwidth,viewport=70 375 350 
665,clip]{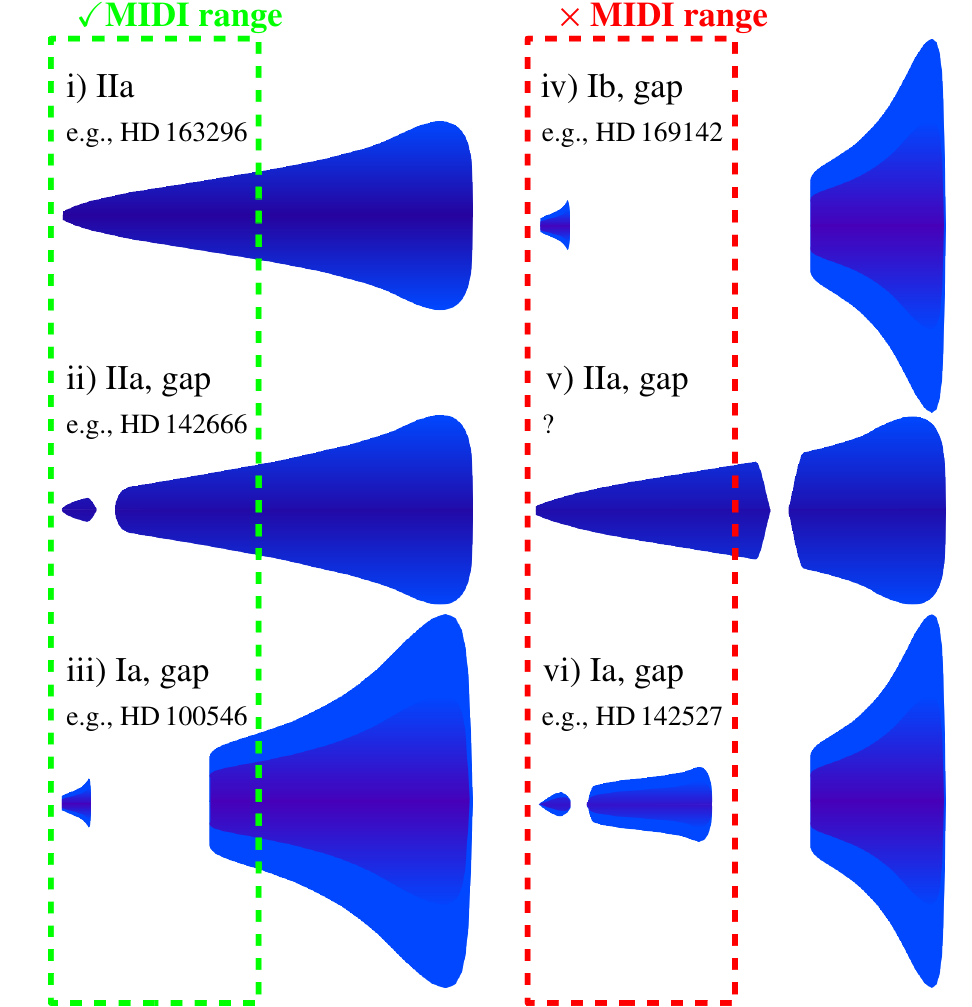}
 \caption{Comparison between the spatial range probed by MIDI (in terms of its spatial resolution and the temperature domain) and the possible, observed disk structures. We distinguish between group I and group II disks, with or without silicate feature (a vs.~b). The green and red colors indicate whether MIDI provides the spatial information to fully characterize the presence and onset (inward) of possible disk gaps.}\label{fig:stages}
\end{figure}

\subsection{The observational picture of Herbig Ae disks}
With MIDI we observe in the $8-13$ micron atmospheric window ($N$ band) and probe mainly warm dusty material at $T\gtrsim200$\,K. This limits the radial range that we see to several tens of au from the star, because at larger radii the dust is not warm enough to emit significantly in this spectral region.\footnote{The radial range seen in the 10-$\mu$m spectral region depends on the stellar luminosity (scaling with $L_\star^{1/2}$). These considerations only apply to dust that is in thermal equilibrium with the radiation field, such as typical silicate dust. In contrast, particles that are stochastically heated, such as PAHs, can be observed to much greater distances of up to several 100 au in some HAe stars. Such particles are small enough to reach temperatures of several 100\,K upon absorption of a single optical/UV photon, making them ``light up'' briefly at mid-infrared wavelengths.} The range of spatial scales that is sampled by our observations and in which we can directly detect the signature of a disk gap is 1 to 15\,au in the best cases, but depends on the UV coverage.

Figure \ref{fig:stages} gives an overview of possible disk geometries for Herbig Ae stars. We distinguish between the following cases:
\begin{enumerate}
 \item Groups I and II \citep{2001A&A...365..476M}. These indicate the amount of far-infrared excess emission and are thought to physically correspond to objects with flared (group I) or flat (group II) outer-disk geometries.
 \item Subgroups a and b \citep{2001A&A...365..476M}, indicating whether emission features from silicate dust are seen (a) or not (b). Physically this corresponds to a configuration where small ($\lesssim$ few micron) silicate grains are present at the disk surface in the right temperature range to emit in the $N$ band (subgroup a), or not (subgroup b). Following the \citet{2013A&A...555A..64M} picture, group Ib objects have very large gaps in the region which otherwise dominates the silicate emission. Transitional group Ia objects still have strong silicate emission, which is dominated by either the outer disk (Ia in Fig.~\ref{fig:stages}iii) or the inner disk (Ia in Fig.~\ref{fig:stages}vi).
\end{enumerate}
The above classification does not provide a sufficient characterization for fully interpreting our MIDI results. In addition to this commonly used framework, we include a third dimension:
\begin{enumerate}
\setcounter{enumi}{2}
 \item The spatial range of a possible disk gap. There may either be no gap, a gap within the spatial range probed by MIDI, or a gap outside the spatial range probed by MIDI. For the three geometries on the left in Fig.~\ref{fig:stages}, MIDI has the capacity to correctly infer whether a gap is present in the disk and to measure the (inward) onset of the gap(s). For the right three geometries, either the main gap is too extended to be identified with MIDI (alone), or it lies outside the temperature or resolution range that is probed. In the group II sources, with a disk gap, the spatial extent of the main gap is smaller than in group I sources.
\end{enumerate}
The geometries depicted in Fig.~\ref{fig:stages} correspond to models that have been proposed for Herbig Ae stars. Two exceptions can be identified: the group IIa disk with a gap far from the star and the group Ia disk with two gaps (Figs.~\ref{fig:stages}v and \ref{fig:stages}vi, respectively).

The first case, the group IIa source with a gapped outer disk may currently be missing owing to an observational bias: the gap in such objects is difficult to detect because the radial width is too small for the gap to have a significant effect on the overall SED. In addition, group II disks are faint in scattered light (e.g., \citealt{2014A&A...568A..40G}), and the gap lies outside the disk region probed with MIDI and can only be detected with long-baseline sub-mm/mm observations with ALMA or the VLA.

The second case, the geometry for a double-gapped group Ia disk in Fig.~\ref{fig:stages}, requires some additional discussion. Several group Ia disks are characterized by large gaps of tens of au in size. Examples are AB\,Aur, HD\,36112, and HD\,142527 with inner radii of the outer disks of 88\,au, 73\,au, and 130\,au, respectively \citep{2010ApJ...718L.199H,2011ApJ...732...42A, 2011A&A...528A..91V}. The physical extent of these gaps, and the fact that they still show strong 10-$\mu$m silicate emission (i.e., subgroup a), indicates that their 10-$\mu$m emission is coming exclusively from an inner disk, a component that is potentially small and partially depleted of dust (e.g., \citealt{2011ApJ...732...42A}). Mid-infrared interferometry of such a source will point to a compact disk with small sizes and blue colors similar to a strongly settled continuous disk. In other words, the sources are expected in the range of models without gaps in Fig.~\ref{sect:sizecolor}. Instead, sources like AB\,Aur (\#7), HD\,36112 (\#11), and HD\,142527 (\#42) have sizes and colors that agree with models with small inner gaps. A possible interpretation is that these group Ia sources have additional small gaps in the inner disk regions, similar to the newly identified gapped group II sources. The depicted geometry in Fig.~\ref{fig:stages} corresponds to such a geometry. Alternatively, the inner disk geometry might be altered in a different way and contain an extended halo component, as was proposed for HD\,142527 (\citealt{2011A&A...528A..91V}; we note that such a component strongly differs from the compact tenuous halo of our radiative transfer models in Sect.~\ref{sect:radtran}). Recent work by \citet{2015ApJ...798L..44M} suggests that the inner disk of HD\,142527 is actually strongly inclined with respect to the outer disk.

\subsection{Effects of gaps on other disk observables}
Is the formation of a gap at a specific location in the disk the fundamental characteristic that determines the observational appearance of Herbig Ae disks? Forming a gap by, for instance, the formation of a sufficiently massive planet will lower the local surface density and largely isolate the inner disk from the rest of the disk. If this process occurs close to the star, continuous accretion of the inner-disk material can partly drain out the inner disk. A lower amount of hot material will lead to a decrease in near-infrared radiation, an effect that will be visible in the object's SED.

\begin{figure}
 \centering
 \includegraphics[width=0.5\textwidth,viewport=15 5 530 400,clip]{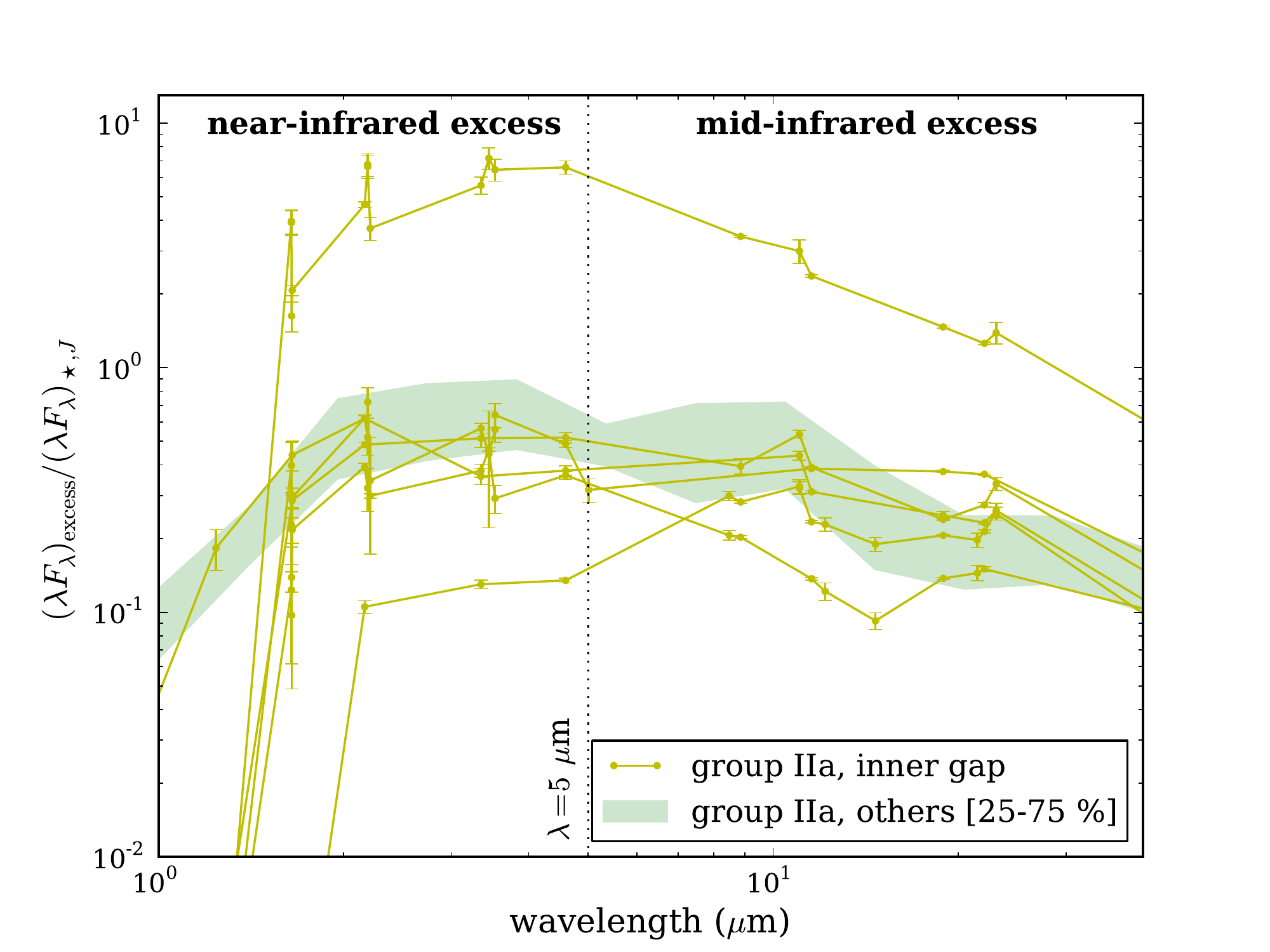}
 \caption{Near- and mid-infrared excess SED of the group IIa stars in the sample. The SEDs are normalized to the stellar flux in the $J$ band ($\lambda=1.24\,\mu$m). The full yellow lines indicate sources with observed or suspected gaps. The shaded green zone show the 25\,\% to 75\,\% percentile of the other objects. Sources with gaps generally have slightly lower near-infrared excesses than gapless sources.}\label{fig:nearIRSED}
\end{figure}

Figure~\ref{fig:nearIRSED} gives an overview of the SEDs of the group II objects in the sample, where the stellar photosphere has been subtracted. In order to compare objects of different luminosities and at different distances, we normalized the fluxes to the photospheric contribution in the $J$ band ($\lambda=1.24\,\mu$m). The aim of the figure is to compare the following classes of sources:
\begin{itemize}
 \item group IIa objects with observed/expected gaps within 15\,au (based on Fig.~\ref{fig:mergedsizecolor}), i.e., sources HD\,72106 (\#22), KK\,Oph (\#56), HD\,142666 (\#41), UX\,Ori (\#10), HD\,144432 (\#45), and AK\,Sco (\#55);
 \item other group IIa objects.
\end{itemize}
Objects with an inner gap tend to have slightly lower near-infrared fluxes than the other group members.\footnote{A notable exception is the upper group IIa source with a (suspected) inner gap. This source is KK\,Oph (\#56). \citet{2013A&A...551A..21K} model its near- and mid-infrared interferometric data with a large envelope with polar cavities, in addition to a (continuous) disk. This geometry deviates from a standard disk, and possibly explains why the source is atypical in our sample.} The mid-infrared fluxes, on the other hand, are similar to the other targets within the group. The lower near-infrared excesses for sources with inner gaps is likely related to the partial accretion (``drain-out'') of the inner disk, as alluded to above.

\citet{2009A&A...502L..17A} find a strong anti-correlation between the 7-$\mu$m excess and the $[30/13.5]$ continuum flux ratio for Herbig Ae/Be stars. In their interpretation, a higher inner disk results in a stronger shadowing of the outer disk, hence a bluer mid-infrared disk color. Conversely, the shadowing effect of a lower inner-disk rim is lower, so these disks appear redder. The relatively low near-infrared excess for disks with an inner gap now provides an alternative explanation for this correlation. In an evolutionary scenario driven by gap formation, objects with a strong near-infrared excess and blue colors have the most primitive inner disks, where a strongly optically thick inner rim casts a shadow on the outer disk. Objects where gap formation in the inner disk has led to a partial drain-out have a lower shadowing effect. These disks are less optically thick, have a more illuminated outer disk, and will appear redder. 

An important diagnostic for the disk structure is the PAH emission. Based on the work of \citet{2010ApJ...718..558A}, \citet{2014A&A...563A..78M} compile a list of 48 Herbig Ae/Be sources and compare their PAH luminosity to the \citet{2001A&A...365..476M} classification. Whereas the group Ib objects all are found to have a relatively strong PAH emission, the PAH contribution in group Ia and IIa objects is not very different. For a similar fraction of sources ($6/13$ for group Ia and $12/28$ for group IIa), PAH emission is not detected, and the intrinsic emission strengths overlap (with a wider spread for the IIa sources). The strong contribution for PAHs in group Ib sources can be attributed to their very large gaps, which lower the continuum emission and increases the irradiation to UV emission \citep{2013A&A...555A..64M,2014A&A...563A..78M}. Group Ia and IIa sources have either smaller or no gaps, which largely excludes this effect. Similarly, the stronger molecular gas emission for group I sources than for group II sources can be related to a more progressed dust clearing within the disks of the former group, resulting in a more efficient heating of the outer layers \citep{2013A&A...559A..84M}.

\subsection{Evolutionary classification for T\,Tauri and Herbig stars}
Herbig Ae stars are the more massive counterparts of the more numerous T\,Tauri stars. Spitzer surveys of clusters (e.g., \citealt{2006AJ....131.1574L,2009ApJ...698....1C, 2010ApJ...718.1200M,2012ApJ...750..157C}) have yielded new insight into the characteristics and evolution of T\,Tauri disks. The SEDs of T\,Tauri stars have been classified in four groups: primordial disks, disks with large inner holes (or gaps), homologously depleted (or anemic/weak-excess) disks with weak overall infrared excesses, and debris disks. The different classes of objects in between primordial and debris disks led to the idea that different evolutionary channels exist toward debris-disk or diskless objects (e.g., \citealt{2006AJ....131.1574L}), governed by the different dispersal processes in protoplanetary disks (e.g., \citealt{2012ApJ...750..157C}). An evolutionary scenario proposed by \citeauthor{2009ApJ...698....1C} (\citeyear{2009ApJ...698....1C}; see also \citealt{2010arXiv1002.1715C}) is that primordial disks evolve into either homologously depleted disks through simultaneous dust clearing at a wide range of disk locations or disks that developed large inner holes (or gaps) as part of an inside-out clearing process. The \citet{2013A&A...555A..64M} scenario can be interpreted as the Herbig Ae equivalent for the above evolutionary scenario, where group II disks are the homologously depleted disks and group I disks the disks with large holes or gaps.

\begin{figure*}
 \centering
 \includegraphics[width=.9\textwidth,viewport=90 500 520 
690,clip]{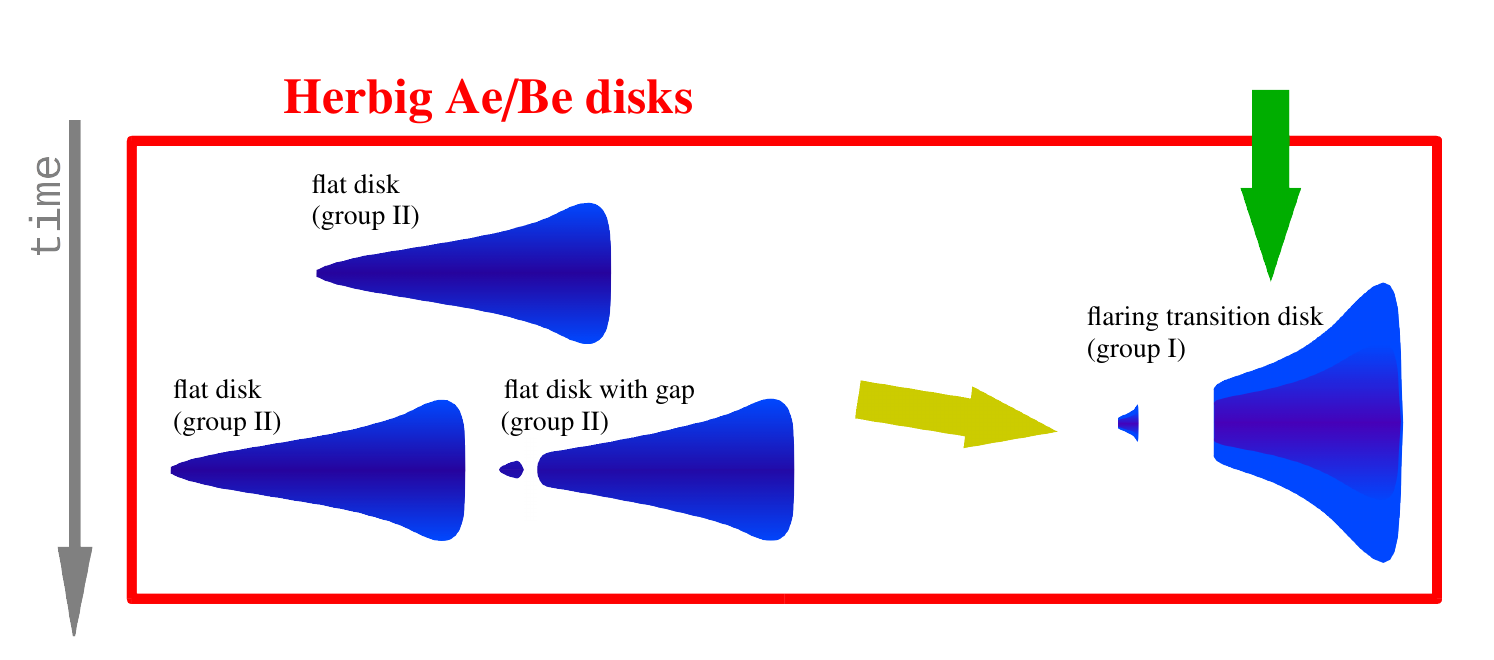}
 \caption{Possible evolutionary connections between Herbig Ae/Be sources. 
Three populations of objects are identified: flat continuous disks (group II sources), flat disks with gaps (group II sources), and flaring transition disks (group I sources). Gapped group II sources are most likely descendants of group II sources without gaps. The arrows indicate two possible evolutionary ideas for the ancestors of transitional group I sources: green arrow = influx from a common ancestor; yellow arrow = influx from group II sources with gaps (cf.~Fig.~\ref{fig:scenarios}).}\label{fig:evolution}
\end{figure*}

The classification and evolutionary scenarios of T\,Tauri stars are essentially based on SEDs, i.e., spatially unresolved data. Even though radiative-transfer modeling may partly lift degeneracies based on simple color criteria (e.g., \citealt{2010ApJ...718.1200M,2011ApJ...732...24C}) and may correctly help to identify large inner holes, the structural information in SEDs is still limited. The group I and group II SED classification for Herbig Ae stars is an excellent example of this. Initial ideas suggested grain growth and settling as the main mechanism to transform group I into group II disks \citep{2001A&A...365..476M,2004A&A...417..159D,2004A&A...421.1075D}, and several works on grain growth endorsed these evolutionary ideas (e.g., \citealt{2004A&A...422..621A,2008ApJ...683..479B,2009A&A...497..379M, 2010ApJ...721..431J}).

Observations that spatially resolve Herbig Ae disks challenged this evolutionary picture. The realization of \citet{2013A&A...555A..64M} that all group I objects may have gaps shows that many ``primordial'' group I disks are in fact in a far-evolved stage of gap formation. Our results provide additional evidence for the gapped nature of group I disks. The presence of large gaps suggests that many disks bear massive, or multiple (e.g., \citealt{2011ApJ...729...47Z,2011ApJ...738..131D}), planets. If all group I objects turn out to have gaps,\footnote{Other recent and upcoming works show additional evidence of gapped group I objects (e.g., \citealt{2014A&A...561A..26M,2015A&A...574A..75V}; Khalafinejad et al., in prep.; Menu et al., in prep.). } it may well be that group I objects indeed form a separate evolutionary channel (\citealt{2013A&A...555A..64M}; see Sect.~\ref{sect:intro}). A possible common ancestor for Herbig Ae disks, like the primordial disk for T\,Tauri stars, has not (yet) been identified.

Our spatially resolved observations of a large sample of Herbig Ae disks, identifying a significant number of gapped group II disks, now provide a second alteration of existing evolutionary scenarios.  The possibility that these objects, in a far stage of dust growth and settling, may also be in an advanced stage of planet formation and disk dissipation is intriguing in its own right. From an evolutionary point of view, the new population of gapped disks may point to new links between the existing classes of objects.

\subsection{Evolutionary implications of gapped group II disks}
A full target-by-target characterization, including the MIDI data and other high-angular-resolution constraints, is an important next step to be taken for understanding the newly identified population of gapped group II disks. In the current picture, the gaps in flat disks are small and are located in the inner few au. Flat disks with small gaps in the outer disk, as in Fig.~\ref{fig:stages}v, have not (yet) been discovered. Based on the limited but qualitatively new information given by the MIDI observations, we now discuss possible evolutionary implications of this newly identified population.

A population of gapped group II disks provides direct evidence that multiple disk-dispersal mechanisms -- grain growth and gap formation -- are contributing together to the evolution of individual disks. The classification of objects based on their spatially unresolved appearance (e.g., the decision tree in Fig.~10 in \citealt{2011ARA&A..49...67W}) is therefore a simplification of the actual dispersal process.

Different ideas for incorporating the new class of objects in current evolutionary scenarios can be thought of. First, the gapped group II objects may be evolved group I objects, with a continued grain growth and settling of the outer disk. Contrary to the gaps discovered in group I disks, gaps in group II disks are small. This scenario would therefore require the collapse of the outer disk and the closing of large gaps to be coeval processes. Though this cannot be excluded, it is not likely to be a common phenomenon. In addition, group Ib objects, which are characterized by extreme gap sizes and a lack of small silicate grains in the inner several tens of au, most probably do not evolve into quasi-continuous disks with a clear silicate feature. Two other evolutionary ideas are the following:
\begin{enumerate}
 \item Even though group II objects have a grain population that experienced significant evolution, gaps may still be developing, yet the gaps are smaller than for the group I objects, and group I and II sources present a parallel evolution.
 \item Group II objects may be the precursors of (some) group I objects. The link between the two could be giant planet formation, leading to the carving of a gap and the dynamical excitation of the disk material.
\end{enumerate}
In both scenarios, group II objects without gaps are the precursors of group II objects with gaps. The main difference between the two ideas is the origin of group I objects. As we depict in Fig.~\ref{fig:evolution} (green arrow), the first scenario assumes that the group I population represents the outcome of an isolated evolutionary channel. The second scenario (yellow arrow if Fig.~\ref{fig:evolution}), on the other hand, assumes at least a partial influx from group II objects.

The first scenario can be embedded naturally in a common-ancestor scenario (for Herbig Ae/Be stars, \citealt{2013A&A...555A..64M}; for T\,Tauri stars, \citealt{2009ApJ...698....1C}). The origin of the different evolutionary outcomes, in terms of the initial conditions, is still not understood. A common ancestor for Herbig Ae/Be stars has not yet been identified. It is therefore not clear whether low- and intermediate-mass objects follow exactly the same evolution. For instance, unlike for evolved T\,Tauri stars, a large number of the intrinsically evolved Herbig Ae objects are accreting (e.g., \citealt{2006A&A...459..837G}), and clear inner holes (as opposed to gaps) seem not to be common. Finally, Herbig Ae disks are often found to be relatively massive (e.g., \citealt{2004A&A...422..621A}). Identifying group II disks with the population of (intrinsically low-mass) homologously depleted T\,Tauri disks is therefore problematic.

The second scenario, which incorporates a possible evolutionary connection between group II and group I objects, is discussed in the next section.

\subsection{Can (some) flat disks evolve in flaring transition disks?}
Age information is very important for verifying different evolutionary ideas. Young stellar clusters, generally having well established ages,\footnote{However, as stated by \citet{2013MNRAS.434..806B}, the ages of star-forming regions may be significantly underestimated. We refer to \citet{2014ApJ...796..127C} for a further discussion of the cluster populations in terms of the previous and new cluster ages.} contain young intermediate-mass objects. It turns out that a large number of these objects in clusters with ages of a few Myr have either debris disks or no disks at all (e.g., \citealt{2006ApJ...638..897S, 2009ApJ...698....1C, 2009AJ....138..703C, 2009ApJ...707..705H}). It is clear that the Herbig Ae stars with massive disks may therefore be a fairly selective population of young stars. For about half of the objects in our sample, an age estimate can be found in \citet{2005A&A...437..189V}. The age values show that many objects in our sample are relatively old ($\gtrsim2\,$Myr), and no clear evidence for a different age distribution for group I and II objects can be inferred. These age estimates of isolated objects have large uncertainties, however. Arguing for different scenarios based on the individual age estimates for the objects is therefore intricate, and the reasoning below is based on more heuristic arguments.

The initial grain growth in disks can be very fast. Within on the order of 0.1\,Myr, the entire submicron grain population of a disk will experience growth and grains with sizes up to 100\,$\mu$m or 1\,cm can be formed (e.g., \citealt{2010A&A...513A..79B}, and references therein). Naturally, by the time the sources have cleared most of the envelope and become optically visible (Lada) class II objects, their disks may have a settled grain population and tend to be ``flat'' (i.e., Meeus group II without gap). Depending on the overall grain growth, disks will be less or more settled (e.g., \citealt{2010ApJ...721..431J}).

The growth from cm-sized pebbles to km-sized planetesimals and finally planets is a more complex process, involving the overcoming of several growth barriers (for a recent review, see \citealt{2014prpl.conf..547J}). The typical timescale for the formation of a massive planet, which is needed for carving out a clear gap in the disk, remains an important unknown quantity. Depending on the exact grain growth models (compact versus fluffy aggregation), the expected timescales for the transition from cm-sized grains to 100-m sized bodies is $10^3$\,yr \citep{2012ApJ...752..106O} to $10^6$\,yr \citep{2012A&A...540A..73W}, at a radial distance of $1\,$au (see also \citealt{1997A&A...325..569S}). Growth to genuine planetesimals possibly even requires mechanisms that lead to a large-scale concentration of particles (e.g., \citealt{2007Natur.448.1022J}), and the characterization of these mechanisms is an ongoing process.

On the time scale of $1-10\,$Myr, objects of $\gtrsim10\,M_\oplus$ can form, which are massive enough to keep hold of the H and He gas (e.g., \citealt{1996Icar..124...62P}). These objects quickly evolve into young gas giant planets of $\gtrsim1\,M_\mathrm{Jup}$, which start clearing a gap in the disk, but this gap is not yet very large and does not significantly affect the SED. Depending on at what radius in the disk the planet forms and opens the gap, we may see it with MIDI if it occurs in the right range (i.e., group II sources with a gap that is detected; Fig.~\ref{fig:stages}ii) or may not see it if the gap lies outside of the range probed by MIDI (i.e., group II sources with a gap that is present but remains undetected; Fig.~\ref{fig:stages}v). 

For a sufficiently massive planet, the gap will become so large (group I) that a significant range of temperatures is ``missing'', yielding the typical ``dip'' in the SED.\footnote{Multiple planets may be required to form large gaps (e.g., \citealt{2011ApJ...729...47Z,2011ApJ...738..131D}).} The massive body dynamically excites the population of planetesimals (e.g.,~\citealt{2011A&A...531A..80K}), leading to a collisional cascade and a renewed production of (sub-)micron sized grains (e.g., \citealt{2003A&A...401..577B}). This increases the opacity of the disk, hence its ability to flare \citep{2004A&A...417..159D}. Furthermore, the direct dynamical interaction of the disk with the massive planet may increase its scale height at the inner edge of the outer disk: energy dissipated in spiral shock waves heat the disk midplane \citep{2002ApJ...565L.109H}, increasing the pressure scale height. Depending on the spatial extent of the gap, we may or may not see the mid-infrared silicate emission. If there is still much material in the right temperature range, we are seeing a group Ia source. In some of these disks, the emission is dominated by material in the inner disk (e.g., HD\,142527), in others by the material in the outer disk (e.g., HD\,100546; see Fig.~\ref{fig:stages}). If the gap becomes so large that the entire region responsible for the mid-infrared silicate emission is empty, we are seeing a group Ib source (e.g., HD\,169142).

The above scenario explains how group II objects may evolve into group I objects via group II objects with gaps. It remains to be shown from a modeling point of view whether such a transformation is possible. It may not be likely or required that all flaring transitional disks are descendants of flat continuous disks. Still, this scenario does not require any unidentified common ancestor, and it naturally explains the occurrence of all the kinds of disks we see, including the absence of group IIb sources. Indeed, for these sources to exist, very large gaps in group II disks would be required (for suppressing the silicate emission), hence very massive or multiple planets. The latter would turn the disk into a group I object long before the gap is large enough to see no silicates.

\section{Summary and conclusions}\label{sect:conclusion}
Linking signatures for the structure of protoplanetary disks to their evolutionary status is one of the primary interests of disk observations. High-angular-resolution observations provide valuable constraints on this discussion. 

Mid-infrared interferometric observations of 64 protoplanetary disks, obtained with the MIDI instrument on the VLTI, were collectively analyzed in this work. The typically low number of observations per target imposes the use of simple disk models. The temperature-gradient model we used leads to size estimates for the disk region emitting in the mid-infrared.

For the total sample, going from low-mass T Tauri objects to massive YSOs, we see a gradual increase in characteristic sizes, but with a considerable spread. Several outliers in the trends are well known for having large disk gaps, implying that the sizes derived using simple disk models provide a diagnostic for the presence of disk gaps.

To quantitatively interpret observed trends, a population of radiative-transfer models was assembled. Limiting the comparison to the Herbig Ae disks, which with 35 representatives form the largest group of objects within the sample, leads to four important findings:
\begin{enumerate}
 \item Most group I objects have sizes and colors that are compatible with having gapped inner disks. This endorses the idea of \citet{2012ApJ...752..143H} and \citet{2013A&A...555A..64M} that many, and maybe all, group I objects have gaps.
 \item Some group II disks, thought to have flat, gapless disks, are indeed compatible with such a configuration.
 \item Several other group II objects have sizes and colors that are incompatible with such a configuration. Instead, they share characteristics with group I objects, which are all thought to have gaps. They also are compatible with radiative-transfer models of transition disks. For some of the mentioned group II disks, it has already been claimed that they have gaps.
 \item Some disks have colors that differ strongly from those of the radiative-transfer models. All these objects are group Ib sources, whose mid-infrared emission (by definition) lacks a silicate feature. We argue that the emission of PAHs and VSGs alters the mid-infrared appearance of these disks and decouples the relation between disk color and (apparent) disk size, usually governed by the dust that is in thermal equilibrium with the radiation field.  
\end{enumerate}
The population of group II disks with gaps makes a strict structural separation between continuous flat disks and flaring transition disks obsolete.

Gaps in massive protoplanetary disks are often associated with the presence of massive planets. It is remarkable that the new population of group II disks also shows gaps. This potentially indicates that these objects are also in an advanced stage of gas-giant planet formation.

From an evolutionary point of view, flat disks with gaps (gapped group II) most likely are descendants of continuous flat disks (group II without gaps). This idea can naturally be embedded in a common-ancestor scenario where flaring transition disks (group I) form a separate evolutionary channel. Gap formation by massive planets may also have a stronger structural impact on the initially settled disk. Gaps growing in size may lead to a group I appearance. A dynamical excitation of large bodies may reintroduce a micron-sized grain population. In addition, planet-disk interaction may dynamically increase the height attained by dust particles. The result of these interactions is a disk that may appear to be group I, with large gaps and a flaring outer disk.

Full grain-growth models for disks, which include the formation of and the interaction with a planetary body, can shed new light on the link between observed disk structures and evolutionary stage.

\begin{acknowledgements}
This work makes use of the Monte Carlo radiative-transfer code MCMax. We wish to thank the main developer Michiel Min for making his code available and providing support for using it. We are also grateful to the anonymous referee for insightful comments, which were valuable for clarifying the evolutionary implications of our results.
\end{acknowledgements}

\bibliography{bibiliography}
\bibliographystyle{aa}

\appendix

\section{Temperature-gradient models: the pole-on approximation}\label{appendix}
The assumption of the temperature-gradient model (Sect.~\ref{sect:model}) that disks are oriented pole-on is incorrect, at least for some disks. In this appendix, we comment further on this approximation and its possible effect on the resulting half-light radii.

The pole-on approximation is equivalent to assuming that the inclination $i$ is zero, an orientation for which the disk's position angle P.A.~is not defined. For the interferometric observation, this orientation has the advantage that the model is independent of the baseline angle. Indeed, the relative angle between the disk's position angle and the baseline angle is what generally plays the role in defining the model orientation. For this reason, the pole-on approximation has generally been used for interferometric surveys with few observations per target (e.g., \citealt{2002ApJ...579..694M,2005ApJ...624..832M}).

\begin{figure}
 \centering
 \includegraphics[width=.5\textwidth]{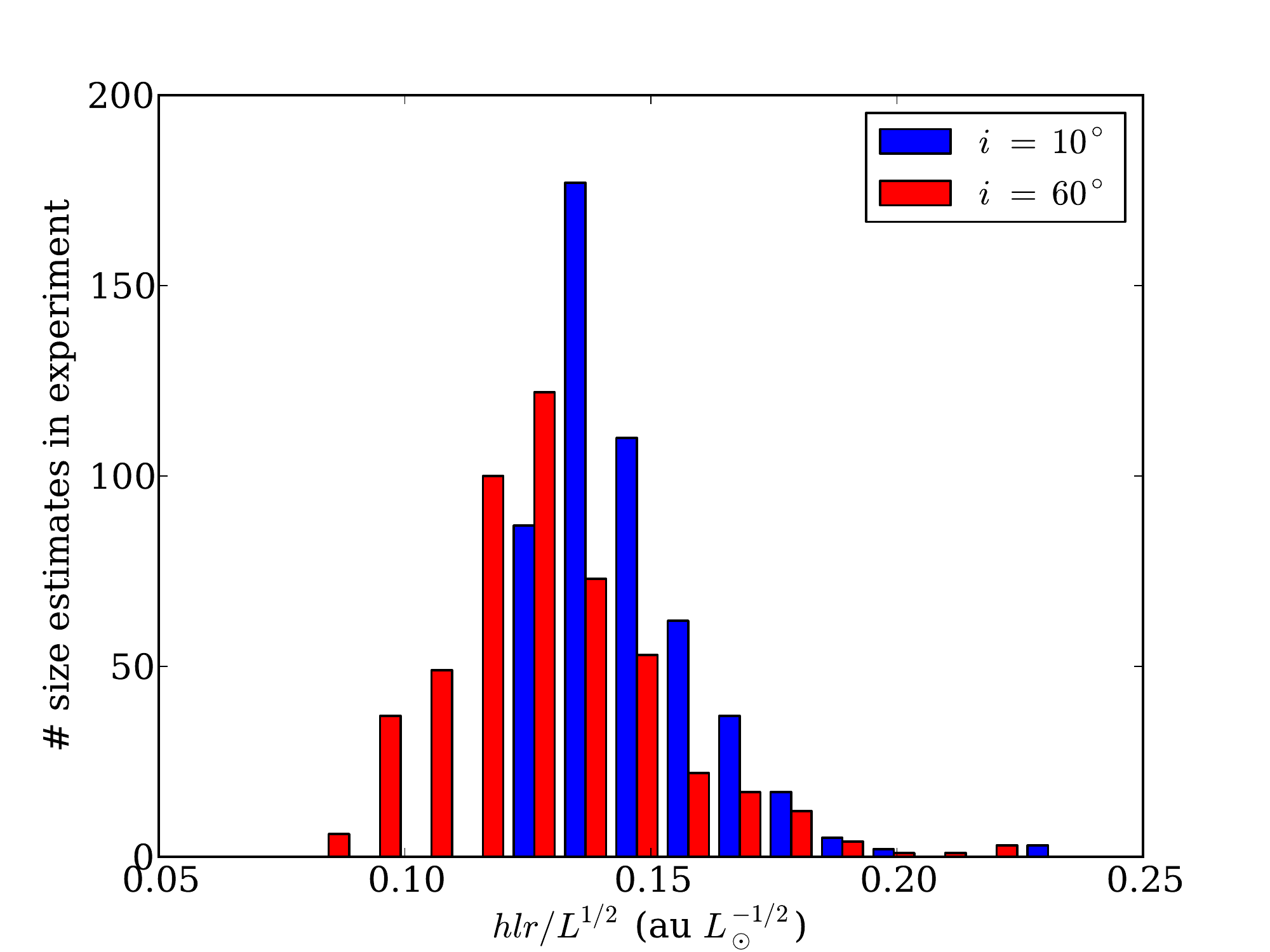}
 \caption{Results of a Monte Carlo simulation for testing the influence of the pole-on approximation of the temperature-gradient models on inclined disks. The blue and red histograms show the distribution of (normalized) half-light radius estimates for a (almost) pole-on disk ($i=10^\circ$) and a strongly inclined disk ($i=60^\circ$), respectively. The median size estimates differ by 10\,\%.}\label{fig:hist_poleonexp}
\end{figure}

Intrinsically, the pole-on approximation is only justified for pole-on or mildly inclined (e.g., $i\lesssim20^\circ$) disks. However, a significant number of the disks in our sample will have a stronger inclination. To justify the use of a pole-on disk geometry for determining the half-light radius of these disks, we perform the following simulation. We take two of the radiative transfer models of Sect.~\ref{sect:radtran} with the same stellar/disk parameters\footnote{Parameter values: $M_\star=2.0\,M_\odot$, $L_\star=21\,L_\odot$, $T_\mathrm{eff}=8500\,$K, $R_\mathrm{in}=R_\mathrm{sub}$, $a_\mathrm{min}=0.01\,\mu$m, $\alpha=10^{-2}$.} but with two different inclinations: $i=10^\circ$ (nearly pole-on) and $i=60^\circ$ (strongly inclined, close to the maximum for a non-obscured central object). For each of the two models, we calculated the half-light radius with the pole-on temperature-gradient model, for a random set of five interferometric observations (i.e., five UV points). This experiment was repeated 500 times, and histograms of the determined half-light radii are shown in Fig.~\ref{fig:hist_poleonexp}. First, the Monte Carlo simulation shows that even for this strong inclination difference, the median half-light radius for both distributions differs by only 10\,\%. Second, the fit of the strongly inclined disks is slightly biased toward underestimating the half-light radii found for the (almost) pole-on disk, and the range of possible size estimates is $20$-$25\,\%$ wider. These minor differences allow us to conclude that the mid-infrared half-light radius of a pole-on temperature gradient model is a robust parameter, even for disks that are strongly inclined.
The conclusions based on Fig.~\ref{fig:mergedsizecolor} (for which the vertical axis is on a logarithmic scale) are thus unaffected by this approximation.

\begin{figure}
 \centering
 \includegraphics[width=.5\textwidth,viewport=40 77 620 350,clip]{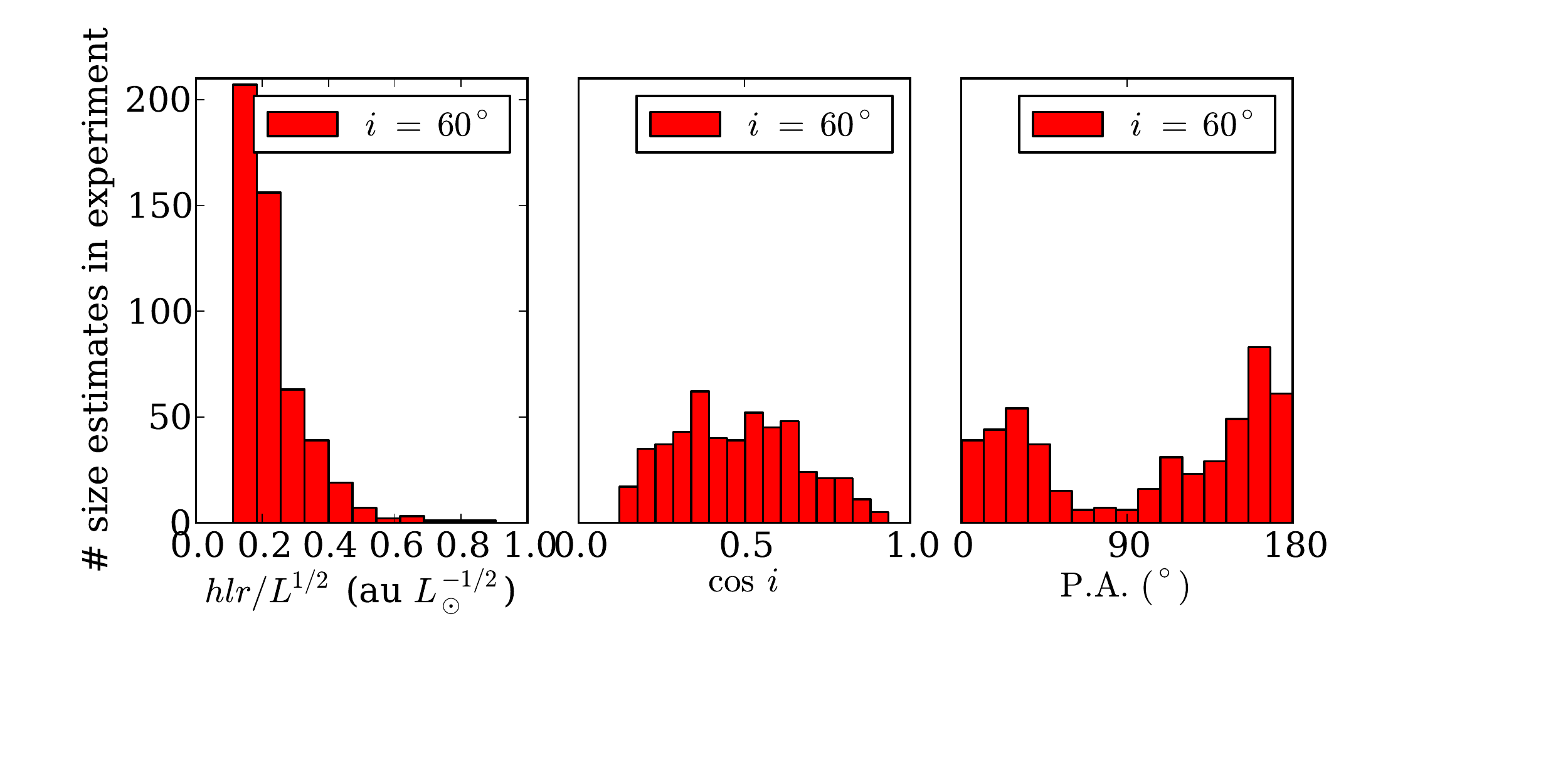}
 \caption{Results for the same Monte Carlo simulation as in Fig.~\ref{fig:hist_poleonexp} (for the radiative-transfer disk with $i=60^\circ$), but with a temperature-gradient model that also includes the inclination $i$ and the position angle P.A.~as free parameters. Clearly, neither the half-light radius, nor $i$ and P.A.~are well constrained. The radiative-transfer disk has $\cos i=0.5$ (i.e., $i=60^\circ$) and $\mathrm{P.A.}=0^\circ/180^\circ$.}\label{fig:hist_poleonexp_inc}
\end{figure}

Two alternatives for this pole-on approximation can be considered, for which we show below that they provide less robust or less confined results. First, it is obviously possible to extend the fit of the temperature-gradient model to include the disk inclination and position angle as fit parameters. We did this experiment for the above radiative-transfer model disk with $i=60^\circ$. In the first histogram in Fig.~\ref{fig:hist_poleonexp_inc}, we see that the inferred half-light radius is much less constrained than under the pole-on approximation in Fig.~\ref{fig:hist_poleonexp}. The two other histograms show the inferred inclination and position angle, neither of which are well constrained. It is clear that the originally robust size parameter (under the pole-on approximation) is not robust when the disk orientation is assumed to be free.

A second option is to fix a non-zero inclination for the temperature-gradient model and determine half-light radii with this inclined geometric model. To avoid biases related to the unknown position angle, the applied model needs to be fit at the full range of position angles ($\textrm{P.A.}=0^\circ$ to $180^\circ/360^\circ$). The result of such a fit is a range of half-light radii (for the varying position angles) rather than a single value. Part of this size range will come from models that are oriented perpendicularly to the actual disk orientation. For strongly inclined disks, these half-light radius estimates will therefore be less precise than when a pole-on model is taken. The result is a less confined size estimate than for the pole-on approximation. The conceptually easier pole-on approximation, which we have shown to be robust (even when disks are strongly inclined), was therefore the preferred approach in this work.

\section{Overview of MIDI observations}\label{appendixB}

\begin{table*}[!h]
\centering
\caption{Overview of observations}\label{table:observations}
\tabcolsep=0.11cm
{\tiny
 \begin{tabular}{l|ccrr|ccrr|ccrr}
\hline\hline name & time & base & $B_p$ & P.A.
& time & base & $B_p$ & P.A.&time & base & $B_p$ & P.A.\\
&(UTC)&&(m)&($^\circ$)\,\,&(UTC)&&(m)&($^\circ$)\,\,&(UTC)&&(m)&($^\circ$)\,\,\\

\hline

LkH$\alpha$ 330&2011-12-11, 04:08&U2U3&40.5&54.1
&2011-12-11, 04:46&U2U3&42.9&51.4
&2011-12-12, 05:24&U1U4&130.1&56.3
\\
&&&&&&&&&&&\\
V892 Tau&2004-12-28, 05:02&U2U3&45.6&46.1
&2004-12-28, 05:05&U2U3&45.7&45.8
&2004-12-30, 04:08&U3U4&48.5&92.0
\\
&2004-12-30, 04:11&U3U4&48.0&91.7
&2009-09-04, 09:43&U1U2&37.2&37.5
&&&&\\
&&&&&&&&&&&\\
RY Tau&2004-11-01, 04:43&U4U2&78.9&96.5
&2004-11-01, 04:57&U4U2&81.0&94.6
&2004-11-04, 07:51&U3U4&48.6&92.0
\\
&2004-12-28, 01:47&U2U3&32.1&54.6
&&&&\\
&&&&&&&&&&&\\
LkCa 15&2011-12-11, 02:31&U2U3&29.9&47.1
&2011-12-11, 03:10&U2U3&33.9&49.7
&2011-12-11, 03:22&U2U3&35.0&50.2
\\
&2011-12-11, 03:29&U2U3&35.8&50.4
&&&&\\
&&&&&&&&&&&\\
DR\,Tau&2004-11-01, 06:47&U4U2&89.0&83.3
&2005-01-01, 02:40&U3U4&61.0&106.2
&2005-01-01, 02:50&U3U4&60.4&105.6
\\
&&&&&&&&&&&\\
GM Aur&2011-12-11, 05:28&U2U3&41.7&52.5
&2011-12-11, 05:41&U2U3&42.5&51.7
&&&&\\
&&&&&&&&&&&\\
AB Aur&2004-12-28, 03:50&U2U3&39.7&54.2
&2004-12-28, 06:00&U2U3&45.9&44.5
&2004-12-30, 05:20&U3U4&42.8&85.4
\\
&2005-01-02, 05:03&U2U4&83.6&66.8
&2005-12-23, 02:48&U1U2&33.9&37.4
&2005-12-24, 03:03&U3U1&69.1&47.5
\\
&2005-12-26, 02:47&U1U4&112.4&75.0
&2005-12-26, 03:00&U1U4&115.3&73.8
&2009-01-21, 01:02&H0G0&29.5&82.7
\\
&2009-01-21, 01:13&H0G0&30.0&81.4
&2009-01-21, 01:45&H0G0&31.1&77.7
&2009-01-21, 01:57&H0G0&31.4&76.3
\\
&2009-01-21, 02:05&H0G0&31.6&75.3
&2009-01-22, 02:16&E0G0&15.9&73.6
&2009-01-22, 03:32&E0H0&47.4&63.6
\\
&2009-01-22, 03:39&E0H0&47.2&62.6
&&&&\\
&&&&&&&&&&&\\
SU Aur&2006-11-10, 06:26&U1U4&120.1&71.5
&&&&\\
&&&&&&&&&&&\\
HD 31648&2007-02-05, 02:08&U2U3&42.8&51.2
&&&&\\
&&&&&&&&&&&\\
UX Ori&2004-12-28, 06:25&U2U3&44.5&44.8
&2004-12-30, 07:02&U3U4&33.0&136.0
&2007-10-25, 07:34&U2U3&44.9&42.1
\\
&2007-10-25, 07:49&U2U3&45.5&43.1
&2007-11-23, 03:09&U1U4&97.7&48.8
&&&&\\
&&&&&&&&&&&\\
HD 36112&2004-12-30, 06:12&U3U4&39.6&89.2
&2005-01-02, 06:18&U2U4&76.7&62.6
&2005-12-23, 03:49&U1U2&39.6&36.1
\\
&2005-12-24, 03:57&U3U1&77.5&44.5
&2005-12-26, 03:54&U1U4&120.8&69.8
&2007-01-03, 02:13&U2U3&31.1&51.3
\\
&2007-01-06, 01:49&U1U3&62.2&38.7
&&&&\\
&&&&&&&&&&&\\
HD 36917&2011-12-11, 07:38&U2U3&45.3&45.9
&2011-12-12, 07:34&U1U4&116.7&62.9
&&&&\\
&&&&&&&&&&&\\
CQ Tau&2005-01-01, 03:38&U3U4&59.8&103.5
&&&&\\
&&&&&&&&&&&\\
V1247 Ori&2011-12-11, 06:57&U2U3&46.6&45.9
&2011-12-12, 06:53&U1U4&126.0&62.7
&2011-12-12, 07:01&U1U4&124.9&62.5
\\
&&&&&&&&&&&\\
HD\,38120&2005-12-23, 01:51&U1U2&49.3&8.4
&2005-12-26, 01:47&U1U4&101.9&49.8
&2005-12-27, 04:27&U2U3&45.9&43.4
\\
&&&&&&&&&&&\\
$\beta$ Pic&2006-12-07, 04:09&U3U4&58.4&94.7
&2006-12-07, 04:43&U3U4&60.2&101.0
&2006-12-07, 05:26&U3U4&61.7&109.0
\\
&2006-12-07, 06:28&U3U4&62.5&120.7
&2007-01-03, 03:44&U2U3&43.7&44.0
&2007-01-04, 04:10&U2U3&42.7&48.5
\\
&2007-01-04, 05:15&U2U3&39.5&57.9
&2007-02-04, 02:48&U2U3&40.9&54.3
&&&&\\
&&&&&&&&&&&\\
HD 45677&2006-10-17, 05:44&H0D0&40.9&41.1
&2006-11-14, 04:26&D0G0&22.8&49.0
&2006-11-14, 07:31&D0G0&31.9&71.0
\\
&2007-10-04, 07:28&E0G0&12.1&52.9
&2007-12-09, 04:06&G1H0&69.9&175.4
&2007-12-11, 08:04&G1H0&71.4&19.7
\\
&2008-01-12, 07:31&E0G0&10.3&78.7
&&&&\\
&&&&&&&&&&&\\
VY Mon&2010-01-13, 03:23&G1D0&67.2&128.4
&2010-01-14, 03:45&H0D0&63.2&74.0
&2010-01-15, 04:19&E0G0&16.0&72.5
\\
&2010-01-16, 04:37&E0H0&47.8&71.5
&2010-01-17, 03:43&H0G0&31.8&73.7
&2010-01-19, 02:56&K0G1&75.1&28.7
\\
&2010-01-19, 05:51&K0G1&89.2&35.8
&2010-01-22, 03:58&K0A0&127.9&72.2
&&&&\\
&&&&&&&&&&&\\
HD 259431&2004-10-30, 08:57&U4U2&89.4&82.1
&2004-11-01, 05:22&U4U2&56.0&90.2
&2004-12-29, 06:41&U2U3&46.5&46.0
\\
&2004-12-30, 02:33&U3U4&59.6&113.7
&2004-12-31, 04:26&U3U4&61.6&107.6
&2005-01-01, 05:43&U3U4&54.7&105.9
\\
&2007-02-08, 05:41&U1U3&102.0&36.8
&2007-03-10, 03:12&U1U2&56.4&35.1
&&&&\\
&&&&&&&&&&&\\
R Mon&2009-01-19, 04:02&H0D0&64.0&72.9
&2009-01-19, 04:27&H0D0&63.8&71.9
&2009-01-19, 05:11&H0D0&62.0&69.6
\\
&2009-01-21, 03:15&H0G0&31.4&74.0
&2009-01-21, 03:26&H0G0&31.7&73.8
&2009-01-22, 03:59&E0H0&48.0&72.5
\\
&2009-01-22, 04:03&E0H0&48.0&72.4
&2009-01-22, 04:28&E0H0&47.6&71.3
&2010-01-13, 04:29&G1D0&61.9&130.9
\\
&2010-01-14, 05:05&H0D0&63.3&71.1
&2010-01-16, 02:24&E0G0&14.1&75.1
&2010-01-16, 05:19&E0H0&46.7&69.9
\\
&2010-01-19, 04:54&K0G1&86.1&35.0
&2010-01-19, 04:58&K0G1&86.4&35.1
&2010-01-21, 02:40&K0A0&120.6&74.8
\\
&2010-01-21, 06:02&A0G1&56.8&116.9
&&&&\\
&&&&&&&&&&&\\
HD 50138&2007-12-09, 03:19&G1H0&68.9&167.8
&2007-12-09, 03:40&G1H0&68.5&169.8
&2007-12-10, 06:33&G1H0&68.4&9.5
\\
&2007-12-12, 05:09&G1D0&71.5&130.0
&2007-12-12, 06:00&G1H0&68.1&6.9
&2007-12-13, 03:10&G1D0&66.6&129.9
\\
&2007-12-26, 07:45&G1D0&58.1&151.3
&2008-11-09, 05:48&E0G0&12.3&60.6
&2008-11-10, 05:43&E0H0&36.7&60.4
\\
&2008-11-10, 05:47&E0H0&37.1&61.0
&2008-12-27, 06:35&E0G0&15.5&74.0
&2008-12-28, 01:39&E0G0&10.0&50.8
\\
&2008-12-28, 02:36&E0H0&36.9&60.8
&2008-12-30, 01:35&H0G0&20.3&51.7
&2009-01-21, 06:01&H0G0&27.6&74.1
\\
&2009-01-21, 06:08&H0G0&27.1&74.0
&2009-03-08, 01:22&E0H0&47.6&73.4
&&&&\\
&&&&&&&&&&&\\
HD 72106&2005-12-30, 07:35&U1U4&124.2&69.5
&2005-12-31, 04:28&U1U4&127.4&40.8
&2006-03-12, 00:53&U2U4&88.2&72.0
\\
&&&&&&&&&&&\\

\end{tabular}
}
\end{table*}

\setcounter{table}{0}

\begin{table*}[!h]
\centering
\caption{continued.}
\tabcolsep=0.11cm
{\tiny
 \begin{tabular}{l|cccc|cccc|ccrr}
\hline\hline name & time & base & $B_p$ & P.A.
& time & base & $B_p$ & P.A.&time & base & $B_p$ & P.A.\\
&(UTC)&&(m)&($^\circ$)\,\,&(UTC)&&(m)&($^\circ$)\,\,&(UTC)&&(m)&($^\circ$)\,\,\\

\hline

HD 87643&2006-02-26, 04:33&D0G0&31.3&76.4
&2006-02-28, 06:22&A0G0&57.1&101.3
&2006-03-01, 01:23&A0G0&63.5&38.2
\\
&2006-03-01, 05:03&A0G0&61.1&85.1
&2006-04-19, 04:34&D0G0&25.4&123.1
&2006-05-23, 02:08&A0G0&51.6&119.9
\\
&2006-05-25, 01:23&A0G0&54.4&110.6
&&&&\\
&&&&&&&&&&&\\
CR Cha&2006-05-19, 04:23&U1U3&57.3&108.4
&2006-05-19, 04:43&U1U3&55.7&114.0
&2007-05-06, 23:38&U3U4&58.7&87.5
\\
&2007-05-06, 23:49&U3U4&58.9&90.1
&2007-06-26, 00:24&U3U4&61.8&145.5
&2007-06-26, 00:40&U3U4&61.9&149.4
\\
&&&&&&&&&&&\\
HD 95881&2004-06-06, 01:35&U3U1&66.4&77.1
&2004-06-07, 01:47&U3U1&64.7&80.6
&&&&\\
&&&&&&&&&&&\\
DI Cha&2006-05-19, 05:23&U1U3&53.9&123.9
&2007-05-07, 00:33&U3U4&59.6&98.7
&2007-05-07, 00:44&U3U4&59.8&101.3
\\
&2007-06-26, 01:09&U3U4&61.9&154.2
&2007-06-26, 01:26&U3U4&61.9&158.2
&2008-05-18, 01:26&U1U3&71.2&64.0
\\
&2008-05-19, 00:46&U1U3&73.3&56.2
&&&&\\
&&&&&&&&&&&\\
HD 97048&2004-12-28, 07:19&U2U3&38.9&28.2
&2004-12-30, 07:55&U3U4&58.5&84.0
&2011-03-20, 04:22&U2U3&36.4&59.4
\\
&2011-03-22, 03:23&U3U4&59.4&95.8
&2011-03-22, 03:27&U3U4&59.5&96.8
&&&&\\
&&&&&&&&&&&\\
HP Cha&2005-05-28, 23:52&U1U2&38.7&44.8
&2005-05-30, 01:01&U3U4&61.2&127.2
&2005-05-30, 01:13&U3U4&61.3&130.1
\\
&2008-04-24, 00:20&U2U4&89.3&65.4
&&&&\\
&&&&&&&&&&&\\
FM Cha&2007-05-07, 01:48&U3U4&60.8&116.4
&2007-05-07, 01:59&U3U4&60.9&119.0
&2008-05-19, 00:00&U1U3&76.6&45.2
\\
&&&&&&&&&&&\\
WW Cha&2005-05-29, 00:59&U1U2&37.0&57.4
&2005-05-29, 23:53&U3U4&60.5&110.9
&2005-05-30, 00:05&U3U4&60.7&113.8
\\
&&&&&&&&&&&\\
CV Cha&2007-06-26, 02:24&U3U4&62.2&170.9
&2007-06-26, 02:35&U3U4&62.2&173.5
&2008-04-21, 04:46&U1U3&65.2&83.3
\\
&&&&&&&&&&&\\
HD 98922&2006-12-30, 08:43&U3U4&59.9&99.5
&2007-01-03, 08:37&U1U4&128.6&56.7
&2007-02-06, 06:02&U1U3&96.6&27.1
\\
&2008-04-08, 07:20&E0G0&11.1&132.9
&2008-05-24, 00:57&H0D0&60.9&85.0
&2008-05-25, 01:52&H0D0&57.1&97.1
\\
&2008-05-25, 02:02&H0D0&56.4&99.2
&2008-05-25, 02:14&H0D0&55.4&101.9
&2008-06-01, 02:11&G1D0&71.2&160.1
\\
&2008-06-01, 02:21&G1D0&71.3&161.9
&2008-06-01, 03:41&G1D0&71.5&177.3
&2009-01-19, 05:38&H0D0&63.1&42.6
\\
&2009-01-21, 04:29&H0G0&30.9&28.4
&2009-01-21, 05:05&H0G0&31.3&36.9
&2009-01-22, 05:13&E0H0&47.1&39.7
\\
&2009-04-20, 05:22&H0G0&25.5&113.2
&2009-04-23, 04:09&E0H0&42.3&99.3
&2009-04-23, 04:57&E0H0&39.2&110.0
\\
&2009-05-24, 23:53&H0D0&63.2&73.2
&2009-05-25, 00:26&H0D0&62.1&79.7
&2009-06-03, 02:51&E0G0&12.3&118.7
\\
&2010-01-17, 07:08&E0G0&16.0&59.6
&2010-01-17, 07:54&E0G0&15.9&68.8
&2010-01-17, 08:44&E0G0&15.6&78.4
\\
&&&&&&&&&&&\\
HD 100453&2004-12-28, 09:01&U2U3&44.3&36.8
&2005-01-02, 09:08&U2U4&89.4&78.8
&2005-12-27, 06:16&U2U3&46.1&7.2
\\
&&&&&&&&&&&\\
HD 100546&2003-06-17, 00:11&U3U1&74.4&60.5
&2004-06-03, 01:24&U3U2&34.8&74.2
&2004-12-28, 08:08&U2U3&41.3&30.8
\\
&2005-12-27, 08:06&U2U3&41.4&29.5
&2006-02-13, 04:01&E0G0&16.0&39.2
&2006-02-16, 05:47&E0G0&15.8&66.6
\\
&2006-02-16, 08:07&E0G0&14.9&99.3
&&&&\\
&&&&&&&&&&&\\
T Cha&2011-03-20, 05:13&U2U3&35.9&61.0
&2011-03-23, 03:23&U2U3&37.7&39.1
&2011-03-24, 03:35&U1U4&121.3&58.1
\\
&&&&&&&&&&&\\
HD 104237&2005-12-27, 06:53&U2U3&39.3&10.5
&2006-03-12, 05:15&U2U3&36.9&53.2
&2006-03-12, 06:28&U2U3&35.2&69.5
\\
&2006-03-12, 06:40&U2U3&34.9&72.3
&2006-05-16, 02:25&U1U3&71.1&63.7
&2006-05-17, 00:54&U1U4&119.9&69.2
\\
&2006-05-18, 02:09&U2U3&35.1&70.6
&2010-05-04, 23:54&K0G1&65.6&16.2
&2010-05-05, 00:13&K0G1&65.3&19.8
\\
&2010-05-05, 00:16&K0G1&65.2&20.6
&&&&\\
&&&&&&&&&&&\\
HD 109085&2007-03-06, 07:49&U1U3&101.9&39.2
&2007-03-08, 05:48&U3U4&62.4&109.0
&2007-03-09, 07:20&U1U3&102.3&38.3
\\
&2007-03-10, 05:27&U2U3&45.6&36.8
&2007-03-10, 05:50&U2U3&46.1&39.1
&2008-03-20, 07:14&U1U2&56.2&34.6
\\
&2010-01-30, 09:34&U2U4&87.2&85.4
&2010-01-30, 09:42&U2U4&86.5&85.9
&2010-02-03, 04:45&U2U3&40.8&8.1
\\
&2010-02-03, 04:54&U2U3&40.9&9.9
&2010-02-03, 08:33&U2U3&46.5&41.5
&2010-02-03, 08:41&U2U3&46.6&42.2
\\
&&&&&&&&&&&\\
DK Cha&2013-05-02, 03:02&U1U3&75.6&47.9
&2013-05-03, 02:33&U1U2&40.1&35.3
&&&&\\
&&&&&&&&&&&\\
HD 135344B&2006-04-15, 05:34&U1U4&130.1&53.7
&2006-05-14, 05:04&U3U4&62.4&115.5
&2006-05-14, 07:39&U3U4&56.3&144.6
\\
&2006-05-16, 03:40&U1U3&101.8&27.7
&2006-06-10, 23:58&U3U4&50.4&87.0
&2006-06-13, 23:42&U1U2&56.5&4.5
\\
&2006-06-14, 02:13&U1U2&55.7&24.8
&2006-07-11, 00:54&U2U4&89.2&83.4
&2006-07-13, 03:37&U2U1&47.7&42.9
\\
&2010-04-25, 03:58&U2U4&84.8&64.3
&2010-04-25, 04:11&U2U4&85.9&66.5
&2010-05-28, 02:02&U1U2&56.4&14.9
\\
&2010-06-28, 00:40&U2U4&88.5&73.4
&2011-03-22, 08:04&U2U4&89.3&82.2
&2011-04-13, 03:47&U2U3&46.4&15.8
\\
&2011-04-14, 05:21&U1U3&102.2&24.3
&2011-04-16, 04:16&U2U3&46.5&22.6
&2011-06-15, 00:53&U3U4&57.8&97.4
\\
&2011-06-15, 01:32&U3U4&60.5&102.9
&2011-06-15, 03:29&U1U4&124.0&69.6
&2011-06-15, 04:07&U1U4&117.9&74.5
\\
&2012-05-10, 06:19&U1U3&94.8&43.1
&2012-06-05, 23:46&U2U4&75.5&46.8
&2012-06-06, 02:43&U2U4&89.4&79.4
\\
&2012-06-06, 04:32&U2U4&83.7&95.9
&&&&\\
&&&&&&&&&&&\\
HD 139614&2010-04-26, 03:16&U3U4&49.8&81.7
&2010-04-26, 04:02&U3U4&54.8&90.0
&2010-04-26, 05:35&U3U4&61.2&105.1
\\
&2011-04-14, 07:58&U1U2&52.4&34.5
&2011-04-18, 03:35&U1U2&56.1&2.1
&2011-04-18, 05:15&U1U2&55.6&16.4
\\
&&&&&&&&&&&\\
HD 142666&2004-06-08, 04:18&U3U1&102.2&36.7
&2004-06-08, 06:50&U3U1&91.0&43.7
&2004-06-08, 07:01&U3U1&89.6&43.7
\\
&&&&&&&&&&&\\
HD 142527&2003-06-14, 00:32&U3U1&102.2&10.9
&2003-06-14, 00:54&U3U1&102.1&14.3
&2005-06-23, 00:58&U1U4&129.6&45.8
\\
&2005-06-23, 02:10&U1U4&129.7&58.0
&2005-06-23, 02:21&U1U4&129.3&59.7
&2005-06-24, 06:33&U3U4&56.1&159.1
\\
&2005-06-27, 05:07&U2U3&36.7&61.2
&2005-06-27, 05:18&U2U3&35.8&62.3
&2005-06-27, 06:18&U2U3&30.0&67.7
\\
&2005-06-27, 23:54&U2U4&83.3&56.4
&2005-06-28, 01:09&U2U4&88.4&70.8
&2005-06-28, 01:20&U2U4&88.8&72.6
\\
&2012-06-06, 00:16&U2U4&77.3&41.9
&&&&\\
&&&&&&&&&&&\\

\end{tabular}
}
\end{table*}

\setcounter{table}{0}

\begin{table*}[!h]
\centering
\caption{continued.}
\tabcolsep=0.11cm
{\tiny
 \begin{tabular}{l|cccc|cccc|ccrr}
\hline\hline name & time & base & $B_p$ & P.A.
& time & base & $B_p$ & P.A.&time & base & $B_p$ & P.A.\\
&(UTC)&&(m)&($^\circ$)\,\,&(UTC)&&(m)&($^\circ$)\,\,&(UTC)&&(m)&($^\circ$)\,\,\\

\hline

HD 142560&2005-05-26, 00:56&U3U4&44.2&78.6
&2005-05-26, 03:43&U3U4&61.1&104.6
&2005-05-26, 07:55&U3U4&55.4&150.1
\\
&2005-06-24, 00:11&U3U4&53.4&90.4
&2005-07-23, 00:54&U2U3&45.5&43.3
&2005-07-23, 04:37&U2U3&31.0&63.1
\\
&2005-08-26, 02:18&U2U4&61.0&125.3
&2005-08-26, 01:55&U2U4&65.3&119.2
&2006-05-15, 02:29&U2U3&46.4&17.2
\\
&2006-05-15, 05:02&U2U3&46.0&40.3
&&&&\\
&&&&&&&&&&&\\
HD 143006&2007-06-26, 03:19&U3U4&61.1&116.7
&&&&\\
&&&&&&&&&&&\\
HD 144432&2003-06-17, 02:15&U3U1&102.3&26.2
&2003-06-17, 04:46&U3U1&98.6&41.1
&2003-06-17, 04:51&U3U1&98.2&41.5
\\
&2003-06-17, 05:07&U3U1&97.0&42.5
&2004-02-11, 08:52&U3U2&45.1&18.5
&2004-02-12, 09:56&U3U2&46.1&29.3
\\
&2004-04-11, 05:55&U3U2&46.0&28.1
&2004-04-12, 06:16&U3U2&46.3&31.6
&2006-03-12, 08:13&U2U3&46.3&30.6
\\
&2006-05-16, 08:41&U1U3&86.2&46.1
&2006-05-16, 08:51&U1U3&84.6&46.3
&2006-07-11, 05:19&U1U3&83.2&46.3
\\
&&&&&&&&&&&\\
HD 144668&2004-04-10, 02:44&U3U2&46.3&173.4
&2004-04-10, 05:23&U3U2&46.6&21.6
&2004-04-11, 05:16&U3U2&46.6&21.1
\\
&2004-04-11, 09:09&U3U2&42.3&52.4
&2004-04-12, 09:46&U3U2&39.8&56.4
&2004-06-28, 02:06&U3U1&100.8&30.9
\\
&2004-06-28, 05:22&U3U1&83.3&50.6
&2004-06-28, 00:15&U3U1&102.4&15.4
&2004-06-28, 04:26&U3U1&90.9&46.1
\\
&2004-09-30, 00:02&U4U2&63.5&123.8
&2006-03-12, 09:39&U2U3&45.6&42.1
&2006-03-12, 09:59&U2U3&45.1&44.6
\\
&2006-07-10, 23:37&U1U4&127.9&41.8
&2009-06-30, 04:09&K0A0&114.8&88.6
&2009-06-30, 04:20&K0A0&112.6&90.3
\\
&2009-07-01, 00:55&K0A0&126.4&58.9
&2009-07-01, 01:42&K0A0&128.0&66.7
&2009-07-01, 01:44&K0A0&128.0&67.2
\\
&2009-07-01, 01:47&K0A0&128.0&67.6
&2009-07-01, 01:50&K0A0&128.0&68.0
&2009-07-01, 05:42&A0G1&83.4&154.1
\\
&2009-07-01, 05:54&A0G1&82.8&156.8
&2009-07-01, 05:58&A0G1&82.7&157.7
&2009-07-01, 06:01&A0G1&82.5&158.6
\\
&2009-07-02, 02:18&K0G1&87.9&27.8
&2009-07-02, 02:22&K0G1&87.7&28.2
&2009-07-04, 00:10&E0H0&46.5&53.0
\\
&2009-07-04, 00:14&E0H0&46.6&53.7
&2009-07-06, 00:30&E0H0&47.3&58.1
&2009-07-06, 00:34&E0H0&47.4&58.8
\\
&2009-07-06, 03:31&E0H0&44.1&86.3
&2009-07-06, 03:35&E0H0&43.8&86.9
&2009-07-07, 03:53&H0G0&28.2&90.3
\\
&2009-07-07, 03:56&H0G0&28.0&90.9
&&&&\\
&&&&&&&&&&&\\
V2246 Oph&2012-05-10, 07:59&U1U3&97.8&42.1
&2012-06-06, 00:54&U2U4&67.6&53.7
&2012-06-06, 01:01&U2U4&68.8&55.1
\\
&&&&&&&&&&&\\
HBC 639&2005-04-18, 04:46&U2U4&74.5&61.2
&2005-04-18, 04:50&U2U4&75.2&61.9
&2005-04-19, 03:53&U2U4&65.3&51.0
\\
&2005-04-19, 04:03&U2U4&67.2&53.3
&2005-08-26, 00:20&U3U4&59.3&121.6
&&&&\\
&&&&&&&&&&&\\
Elias 2-24&2006-05-19, 08:31&U1U3&90.9&44.4
&2007-05-07, 04:32&U3U4&58.3&101.9
&2007-05-07, 04:39&U3U4&58.8&102.5
\\
&2007-05-07, 05:19&U3U4&61.4&106.0
&2007-05-07, 05:30&U3U4&61.8&107.0
&2008-04-21, 08:15&U1U3&101.3&38.3
\\
&2008-05-18, 03:56&U1U3&101.1&21.8
&2008-05-18, 04:38&U1U3&102.0&27.4
&2008-05-19, 05:51&U1U3&102.2&35.5
\\
&&&&&&&&&&&\\
Elias 2-28&2012-06-06, 03:14&U2U4&87.5&75.4
&&&&\\
&&&&&&&&&&&\\
Elias 2-30&2012-06-06, 01:30&U2U4&74.0&60.7
&2012-06-06, 01:41&U2U4&75.7&62.5
&2012-06-06, 01:47&U2U4&76.8&63.6
\\
&&&&&&&&&&&\\
V2129 Oph&2011-05-14, 06:03&U3U4&62.3&113.0
&&&&\\
&&&&&&&&&&&\\
V2062 Oph&2007-06-26, 04:11&U3U4&60.2&119.6
&2008-05-17, 04:20&U1U3&101.5&23.9
&2011-04-13, 04:56&U2U3&44.1&15.4
\\
&2011-04-13, 08:38&U2U3&46.2&44.7
&2011-04-14, 09:35&U2U3&44.4&48.9
&2011-04-16, 07:56&U3U4&62.3&112.9
\\
&2011-04-17, 05:13&U3U4&53.7&98.2
&&&&\\
&&&&&&&&&&&\\
HD\,150193&2007-06-03, 03:43&U1U4&128.0&54.3
&2008-03-22, 05:24&U2U4&59.1&43.1
&2008-03-23, 05:30&U2U4&60.9&45.8
\\
&2008-03-24, 07:36&U1U4&123.6&48.3
&2008-03-24, 07:52&U1U4&125.4&50.6
&2008-06-23, 06:21&U2U3&40.1&51.5
\\
&2013-04-29, 06:20&U3U4&62.1&107.8
&2013-04-30, 05:06&U1U4&122.9&47.5
&2013-04-30, 05:25&U1U4&125.1&50.3
\\
&2013-04-30, 06:04&U1U4&128.7&55.5
&2013-05-02, 05:01&U1U3&100.6&20.1
&2013-05-03, 04:20&U1U2&55.7&9.0
\\
&2014-04-16, 06:50&U3U4&61.2&105.7
&&&&\\
&&&&&&&&&&&\\
AK Sco&2005-05-29, 03:27&U1U2&56.4&13.6
&2005-05-29, 07:14&U3U1&91.8&45.6
&2005-05-30, 02:48&U3U4&53.5&91.1
\\
&2005-05-30, 03:04&U3U4&55.2&93.4
&2005-05-30, 08:44&U3U4&54.6&151.6
&&&&\\
&&&&&&&&&&&\\
KK Oph&2003-06-17, 01:19&U3U1&100.3&8.8
&2003-06-17, 03:19&U3U1&102.2&26.2
&2003-06-17, 03:40&U3U1&102.4&28.8
\\
&2003-06-17, 05:55&U3U1&98.4&41.4
&&&&\\
&&&&&&&&&&&\\
51 Oph&2003-06-15, 03:25&U3U1&101.3&23.3
&2003-06-15, 03:35&U3U1&101.5&24.7
&2003-06-15, 06:54&U3U1&97.3&42.4
\\
&2003-06-15, 08:23&U3U1&86.4&44.5
&2003-06-16, 00:01&U3U1&98.9&171.8
&2003-06-16, 00:09&U3U1&98.8&173.2
\\
&2003-06-16, 02:17&U3U1&99.6&14.1
&2006-05-16, 05:23&U1U3&101.3&23.5
&2006-07-10, 03:55&U2U4&87.4&86.7
\\
&&&&&&&&&&&\\
HD 163296&2003-06-14, 03:13&U3U1&99.4&17.7
&2009-07-07, 07:56&H0G0&16.8&92.7
&2009-07-07, 07:59&H0G0&16.4&93.1
\\
&2009-08-14, 03:12&E0G0&14.2&81.3
&2009-08-14, 03:53&E0G0&12.8&84.4
&2009-08-15, 00:16&H0G0&31.4&66.1
\\
&2009-08-15, 00:31&H0G0&31.7&67.8
&2009-08-15, 02:26&H0G0&30.5&78.1
&2009-08-15, 03:00&H0G0&28.9&80.7
\\
&2009-08-15, 03:12&H0G0&28.2&81.7
&2010-05-05, 05:03&A0G1&74.7&104.2
&2010-05-05, 05:07&A0G1&75.5&104.4
\\
&2010-05-21, 04:51&H0E0&43.7&57.8
&2010-05-21, 05:02&H0E0&44.5&59.5
&2010-05-21, 06:06&H0I1&40.7&146.3
\\
&2010-05-21, 07:05&H0I1&40.6&152.1
&2010-05-21, 07:59&H0I1&40.1&158.7
&2010-05-21, 08:52&I1E0&60.6&122.2
\\
&2010-05-21, 09:03&I1E0&59.3&124.1
&&&&\\
&&&&&&&&&&&\\
HD 169142&2011-04-15, 06:23&U2U3&45.1&11.8
&2011-04-15, 08:59&U2U3&46.6&36.1
&2011-04-15, 09:09&U2U3&46.6&37.4
\\
&2011-04-16, 08:49&U3U4&61.5&106.0
&2011-04-17, 07:52&U3U4&58.2&100.2
&2011-04-17, 08:00&U3U4&58.9&101.1
\\
&2012-06-06, 05:05&U2U4&87.4&73.5
&2012-06-06, 05:13&U2U4&88.0&74.6
&2012-06-06, 05:31&U2U4&88.8&77.0
\\
&&&&&&&&&&&\\

\end{tabular}
}
\end{table*}

\setcounter{table}{0}

\begin{table*}[!h]
\centering
\caption{continued.}
\tabcolsep=0.11cm
{\tiny
 \begin{tabular}{l|cccc|cccc|ccrr}
\hline\hline name & time & base & $B_p$ & P.A.
& time & base & $B_p$ & P.A.&time & base & $B_p$ & P.A.\\
&(UTC)&&(m)&($^\circ$)\,\,&(UTC)&&(m)&($^\circ$)\,\,&(UTC)&&(m)&($^\circ$)\,\,\\

\hline

MWC 297&2006-04-20, 07:55&D0G0&28.9&68.8
&2006-04-20, 09:47&D0G0&32.0&72.8
&2006-04-20, 10:29&D0G0&31.4&73.2
\\
&2006-04-20, 06:02&D0G0&20.1&56.4
&2006-04-21, 09:09&D0G0&31.7&72.1
&2006-04-22, 09:59&D0G0&31.9&73.1
\\
&2006-05-23, 05:06&A0G0&52.6&66.0
&2006-05-23, 06:52&A0G0&63.1&71.8
&2006-05-25, 05:12&A0G0&54.5&67.1
\\
&2006-05-25, 07:20&A0G0&64.0&72.7
&2007-04-12, 07:52&E0G0&13.3&66.3
&2007-04-12, 07:59&E0G0&13.5&66.8
\\
&2007-04-12, 08:22&E0G0&14.3&68.5
&2007-04-17, 06:50&H0G0&23.2&61.9
&2007-04-18, 08:36&E0G0&15.2&70.5
\\
&2007-04-19, 08:00&E0H0&43.3&68.9
&2007-04-24, 06:54&G1D0&70.5&129.5
&2007-04-24, 07:06&G1D0&70.9&129.4
\\
&2007-04-24, 07:12&G1D0&71.1&129.4
&2007-04-25, 07:13&H0D0&54.8&67.3
&2007-04-25, 09:42&G1D0&66.1&135.9
\\
&2007-04-25, 09:54&G1D0&65.0&137.1
&2007-05-08, 06:41&G1D0&71.5&129.5
&2007-05-10, 05:45&G1H0&66.8&172.9
\\
&&&&&&&&&&&\\
MWC 300&2009-04-16, 07:02&E0H0&36.5&61.2
&2009-04-20, 05:56&H0G0&20.1&52.7
&2009-05-03, 08:34&G1D0&69.2&133.6
\\
&2009-05-04, 06:21&G1H0&67.6&174.3
&2009-05-04, 08:52&G1H0&68.4&11.2
&2009-06-30, 05:24&K0A0&127.4&73.2
\\
&2009-06-30, 05:35&K0A0&126.8&73.4
&2009-06-30, 06:06&K0A0&123.5&73.8
&2009-07-01, 04:25&A0G1&90.0&114.5
\\
&2009-07-04, 05:56&E0H0&46.0&73.8
&2009-07-04, 07:26&E0H0&37.7&72.9
&2009-07-04, 07:34&E0H0&36.7&72.6
\\
&2009-07-05, 04:28&E0G0&16.0&72.3
&2009-07-06, 05:57&E0H0&45.4&73.9
&2009-07-06, 06:01&E0H0&45.2&73.9
\\
&2009-07-07, 05:01&H0G0&31.8&73.3
&2009-07-07, 05:05&H0G0&31.7&73.4
&2010-05-04, 09:34&K0A0&125.6&73.6
\\
&2010-05-04, 09:40&K0A0&124.9&73.7
&2010-05-20, 05:49&H0I1&40.7&142.8
&2012-05-01, 07:34&D0B2&31.3&21.0
\\
&2012-05-02, 05:59&B2A1&10.2&113.2
&2012-05-02, 09:55&D0B2&33.8&33.3
&&&&\\
&&&&&&&&&&&\\
R CrA&2004-07-09, 09:10&U3U2&27.4&63.5
&2004-07-28, 02:04&U3U2&46.6&30.5
&2004-07-29, 23:49&U3U2&46.2&9.4
\\
&2004-07-30, 03:32&U3U2&45.6&43.3
&2004-07-30, 05:33&U3U2&40.1&55.5
&2004-07-30, 05:40&U3U2&39.6&56.1
\\
&2005-06-23, 07:33&U1U4&115.3&76.1
&2005-06-25, 07:29&U1U4&114.6&76.6
&2005-06-25, 07:39&U1U4&112.5&77.8
\\
&2005-06-26, 05:57&U1U4&127.5&65.3
&2005-06-28, 04:40&U3U4&61.2&105.0
&2005-06-28, 07:40&U3U4&58.8&134.5
\\
&2005-07-21, 01:29&U3U4&53.3&90.8
&2005-07-21, 02:26&U3U4&58.7&98.9
&2005-07-21, 04:35&U3U4&62.3&117.6
\\
&2005-09-18, 02:10&U1U4&111.2&78.5
&2009-07-01, 06:55&A0G1&88.3&133.1
&2009-07-01, 06:59&A0G1&88.1&133.7
\\
&2009-07-04, 08:36&E0H0&32.3&106.1
&2009-07-04, 08:43&E0H0&31.4&107.7
&2009-07-05, 01:21&E0G0&14.0&32.5
\\
&2009-07-06, 04:43&E0H0&47.9&70.9
&2009-07-06, 07:38&E0H0&37.8&96.8
&&&&\\
&&&&&&&&&&&\\
T CrA&2004-08-01, 03:47&U3U2&45.0&45.9
&&&&\\
&&&&&&&&&&&\\
HD 179218&2003-06-16, 03:17&U3U1&59.7&12.9
&2003-06-16, 03:31&U3U1&61.0&16.6
&2003-06-16, 06:09&U3U1&84.7&39.4
\\
&2004-04-10, 07:43&U3U2&25.4&27.0
&2004-04-10, 08:26&U3U2&29.1&36.4
&2006-05-15, 06:11&U2U3&29.1&36.5
\\
&2006-05-16, 08:01&U1U3&83.2&38.8
&2006-05-16, 08:12&U1U3&85.0&39.5
&2006-05-17, 05:39&U3U4&58.2&118.0
\\
&2006-06-11, 06:42&U3U4&60.3&105.7
&2006-06-14, 06:53&U1U2&48.2&34.6
&2006-06-14, 08:20&U1U2&53.4&36.8
\\
&2006-07-09, 06:58&U3U4&44.4&101.1
&2006-07-13, 06:14&U1U2&52.8&36.8
&2006-07-13, 06:27&U1U2&53.5&36.8
\\
&2009-08-14, 00:09&E0G0&11.6&81.0
&2009-08-14, 00:47&E0G0&13.2&79.7
&2009-08-14, 01:20&E0G0&14.3&78.3
\\
&2009-08-14, 01:51&E0G0&15.1&76.9
&2009-08-14, 02:03&E0G0&15.3&76.3
&2009-08-14, 02:36&E0G0&15.8&74.5
\\
&2009-08-14, 04:24&E0G0&15.4&66.5
&2009-08-15, 01:36&H0G0&29.7&77.5
&2009-08-15, 02:10&H0G0&31.1&75.7
\\
&&&&&&&&&&&\\

\hline

\end{tabular}
}
\end{table*}

\end{document}